\documentclass[twocolumn]{aastex701}

\begin{document}
\maxdeadcycles=200

\newcommand{\LyA}{Lyman-$\alpha$}
\newcommand{\Mgii}{\ion{Mg}{2}}
\newcommand{\EWFL}{W$_0^{\lambda 2796}$}
\newcommand{\kms}{${\rm km\,s}^{-1}$}
\newcommand{\voff}{${v}_{\rm off}$}
\newcommand{\avgebv}{$\overline{\textrm{E(B-V)}}$}

\newcommand{\LGN}[1]{\textcolor{blue}{\bf #1}}1
\newcommand{\citehere}{\textcolor{red}{\bf Citation Needed}} 
\newcommand{\ADM}[1]{\textcolor{orange}{\bf ADM: #1}}1

\title{The Composite Spectrum of QSO Absorption Line Systems in DESI DR2}

\author[0000-0002-5166-8671]{Lucas Napolitano}
\email{lucas.napolitano@noirlab.edu}
\affiliation{Department of Physics \& Astronomy, University of Wyoming, 1000 E. University, Dept. 3905, Laramie, WY 82071, USA}
\affiliation{NSF NOIRLab, 950 N. Cherry Ave., Tucson, AZ 85719, USA}

\author{Adam D.\ Myers}
\email{amyers14@uwyo.edu}
\affiliation{Department of Physics \& Astronomy, University of Wyoming, 1000 E. University, Dept. 3905, Laramie, WY 82071, USA}

\author{Adam Tedeschi}
\email{atedesch@uwyo.edu}
\affiliation{Department of Physics \& Astronomy, University of Wyoming, 1000 E. University, Dept. 3905, Laramie, WY 82071, USA}

\author[0000-0003-2923-1585]{Abhijeet Anand}
\email{abhijeetanand2011@gmail.com}
\affiliation{Lawrence Berkeley National Laboratory, 1 Cyclotron Road, Berkeley, CA 94720, USA}

\author[0000-0002-9136-9609]{Hiram K. Herrera-Alcantar}
\email{herreraa@iap.fr}
\affiliation{Institut d'Astrophysique de Paris. 98 bis boulevard Arago. 75014 Paris, France}
\affiliation{IRFU, CEA, Universit\'{e} Paris-Saclay, F-91191 Gif-sur-Yvette, France}

\author{Jessica Aguilar}
\email{jaguilar@lbl.gov}
\affiliation{Lawrence Berkeley National Laboratory, 1 Cyclotron Road, Berkeley, CA 94720, USA}

\author[0000-0001-6098-7247]{Steven Ahlen}
\email{ahlen@bu.edu}
\affiliation{Physics Dept., Boston University, 590 Commonwealth Avenue, Boston, MA 02215, USA}

\author[0000-0003-4162-6619]{Stephen Bailey}
\email{stephenbailey@lbl.gov}
\affiliation{Lawrence Berkeley National Laboratory, 1 Cyclotron Road, Berkeley, CA 94720, USA}

\author[0000-0001-5537-4710]{Segev BenZvi}
\email{sbenzvi@ur.rochester.edu}
\affiliation{Department of Physics \& Astronomy, University of Rochester, 206 Bausch and Lomb Hall, P.O. Box 270171, Rochester, NY 14627-0171, USA}

\author[0000-0001-9712-0006]{Davide Bianchi}
\email{davide.bianchi1@unimi.it}
\affiliation{Dipartimento di Fisica ``Aldo Pontremoli'', Universit\`a degli Studi di Milano, Via Celoria 16, I-20133 Milano, Italy}
\affiliation{INAF-Osservatorio Astronomico di Brera, Via Brera 28, 20122 Milano, Italy}

\author{David Brooks}
\email{david.brooks@ucl.ac.uk}
\affiliation{Department of Physics \& Astronomy, University College London, Gower Street, London, WC1E 6BT, UK}

\author{Todd Claybaugh}
\email{tmclaybaugh@lbl.gov}
\affiliation{Lawrence Berkeley National Laboratory, 1 Cyclotron Road, Berkeley, CA 94720, USA}

\author[0000-0002-2169-0595]{Andrei Cuceu}
\email{acuceu@lbl.gov}
\affiliation{Lawrence Berkeley National Laboratory, 1 Cyclotron Road, Berkeley, CA 94720, USA}

\author[0000-0002-1769-1640]{Axel de la Macorra}
\email{macorra@fisica.unam.mx}
\affiliation{Instituto de F\'{\i}sica, Universidad Nacional Aut\'{o}noma de M\'{e}xico,  Cd. de M\'{e}xico  C.P. 04510,  M\'{e}xico}

\author[0000-0002-4928-4003]{Arjun Dey}
\email{arjun.dey@noirlab.edu}
\affiliation{NSF NOIRLab, 950 N. Cherry Ave., Tucson, AZ 85719, USA}

\author[0000-0002-5665-7912]{Biprateep Dey}
\email{b.dey@utoronto.ca}
\affiliation{Department of Astronomy \& Astrophysics, University of Toronto, Toronto, ON M5S 3H4, Canada}
\affiliation{Department of Physics \& Astronomy and Pittsburgh Particle Physics, Astrophysics, and Cosmology Center (PITT PACC), University of Pittsburgh, 3941 O'Hara Street, Pittsburgh, PA 15260, USA}

\author{Peter Doel}
\email{apd@star.ucl.ac.uk}
\affiliation{Department of Physics \& Astronomy, University College London, Gower Street, London, WC1E 6BT, UK}

\author[0000-0002-3033-7312]{Andreu Font-Ribera}
\email{afont@ifae.es}
\affiliation{Institut de F\'{i}sica d’Altes Energies (IFAE), The Barcelona Institute of Science and Technology, Campus UAB, 08193 Bellaterra Barcelona, Spain}

\author[0000-0002-2890-3725]{Jaime E. Forero-Romero}
\email{je.forero@uniandes.edu.co}
\affiliation{Departamento de F\'isica, Universidad de los Andes, Cra. 1 No. 18A-10, Edificio Ip, CP 111711, Bogot\'a, Colombia}
\affiliation{Observatorio Astron\'omico, Universidad de los Andes, Cra. 1 No. 18A-10, Edificio H, CP 111711 Bogot\'a, Colombia}

\author{Enrique Gaztañaga}
\email{gaztanaga@gmail.com}
\affiliation{Institut d'Estudis Espacials de Catalunya (IEEC), 08034 Barcelona, Spain}
\affiliation{Institute of Cosmology and Gravitation, University of Portsmouth, Dennis Sciama Building, Portsmouth, PO1 3FX, UK}
\affiliation{Institute of Space Sciences, ICE-CSIC, Campus UAB, Carrer de Can Magrans s/n, 08913 Bellaterra, Barcelona, Spain}

\author[0000-0003-3142-233X]{Satya Gontcho A Gontcho}
\email{satya@virginia.edu}
\affiliation{Lawrence Berkeley National Laboratory, 1 Cyclotron Road, Berkeley, CA 94720, USA}
\affiliation{University of Virginia, Department of Astronomy, Charlottesville, VA 22904, USA}

\author{Gaston Gutierrez}
\email{gaston@fnal.gov}
\affiliation{Fermi National Accelerator Laboratory, PO Box 500, Batavia, IL 60510, USA}

\author[0000-0001-9822-6793]{Julien Guy}
\email{jguy@lbl.gov}
\affiliation{Lawrence Berkeley National Laboratory, 1 Cyclotron Road, Berkeley, CA 94720, USA}

\author[0000-0003-0201-5241]{Dick Joyce}
\email{richard.joyce@noirlab.edu}
\affiliation{NSF NOIRLab, 950 N. Cherry Ave., Tucson, AZ 85719, USA}

\author[0000-0001-6356-7424]{Anthony Kremin}
\email{akremin@lbl.gov}
\affiliation{Lawrence Berkeley National Laboratory, 1 Cyclotron Road, Berkeley, CA 94720, USA}

\author[0000-0003-1838-8528]{Martin Landriau}
\email{mlandriau@lbl.gov}
\affiliation{Lawrence Berkeley National Laboratory, 1 Cyclotron Road, Berkeley, CA 94720, USA}

\author[0000-0001-7178-8868]{Laurent Le Guillou}
\email{llg@lpnhe.in2p3.fr}
\affiliation{Sorbonne Universit\'{e}, CNRS/IN2P3, Laboratoire de Physique Nucl\'{e}aire et de Hautes Energies (LPNHE), FR-75005 Paris, France}

\author[0000-0003-4962-8934]{Marc Manera}
\email{mmanera@ifae.es}
\affiliation{Departament de F\'{i}sica, Serra H\'{u}nter, Universitat Aut\`{o}noma de Barcelona, 08193 Bellaterra (Barcelona), Spain}

\author[0000-0002-1125-7384]{Aaron Meisner}
\email{aaron.meisner@noirlab.edu}
\affiliation{NSF NOIRLab, 950 N. Cherry Ave., Tucson, AZ 85719, USA}

\author{Ramon Miquel}
\email{rmiquel@ifae.es}
\affiliation{Instituci\'{o} Catalana de Recerca i Estudis Avan\c{c}ats, Passeig de Llu\'{\i}s Companys, 23, 08010 Barcelona, Spain}
\affiliation{Institut de F\'{i}sica d’Altes Energies (IFAE), The Barcelona Institute of Science and Technology, Campus UAB, 08193 Bellaterra Barcelona, Spain}

\author[0000-0002-2733-4559]{John Moustakas}
\email{jmoustakas@siena.edu}
\affiliation{Department of Physics and Astronomy, Siena College, 515 Loudon Road, Loudonville, NY 12211, USA}

\author[0000-0001-9070-3102]{Seshadri Nadathur}
\email{seshadri.nadathur@port.ac.uk}
\affiliation{Institute of Cosmology and Gravitation, University of Portsmouth, Dennis Sciama Building, Portsmouth, PO1 3FX, UK}

\author[0000-0003-3188-784X]{Nathalie Palanque-Delabrouille}
\email{npalanque-delabrouille@lbl.gov}
\affiliation{IRFU, CEA, Universit\'{e} Paris-Saclay, F-91191 Gif-sur-Yvette, France}
\affiliation{Lawrence Berkeley National Laboratory, 1 Cyclotron Road, Berkeley, CA 94720, USA}

\author[0000-0002-0644-5727]{Will Percival}
\email{will.percival@uwaterloo.ca}
\affiliation{Department of Physics and Astronomy, University of Waterloo, 200 University Ave W, Waterloo, ON N2L 3G1, Canada}
\affiliation{Perimeter Institute for Theoretical Physics, 31 Caroline St. North, Waterloo, ON N2L 2Y5, Canada}
\affiliation{Waterloo Centre for Astrophysics, University of Waterloo, 200 University Ave W, Waterloo, ON N2L 3G1, Canada}

\author[0000-0001-7145-8674]{Francisco Prada}
\email{fprada@iaa.es}
\affiliation{Instituto de Astrof\'{i}sica de Andaluc\'{i}a (CSIC), Glorieta de la Astronom\'{i}a, s/n, E-18008 Granada, Spain}

\author[0000-0001-6979-0125]{Ignasi P\'erez-R\`afols}
\email{ignasi.perez.rafols@upc.edu}
\affiliation{Departament de F\'isica, EEBE, Universitat Polit\`ecnica de Catalunya, c/Eduard Maristany 10, 08930 Barcelona, Spain}

\author{Graziano Rossi}
\email{graziano@sejong.ac.kr}
\affiliation{Department of Physics and Astronomy, Sejong University, Seoul, 143-747, Korea}

\author[0000-0002-9646-8198]{Eusebio Sanchez}
\email{eusebio.sanchez@ciemat.es}
\affiliation{CIEMAT, Avenida Complutense 40, E-28040 Madrid, Spain}

\author{David Schlegel}
\email{djschlegel@lbl.gov}
\affiliation{Lawrence Berkeley National Laboratory, 1 Cyclotron Road, Berkeley, CA 94720, USA}

\author{Michael Schubnell}
\email{schubnel@umich.edu}
\affiliation{Department of Physics, University of Michigan, Ann Arbor, MI 48109, USA}
\affiliation{University of Michigan, Ann Arbor, MI 48109, USA}

\author[0000-0002-3461-0320]{Joesph Harry Silber}
\email{jhsilber@lbl.gov}
\affiliation{Lawrence Berkeley National Laboratory, 1 Cyclotron Road, Berkeley, CA 94720, USA}

\author{David Sprayberry}
\email{david.sprayberry@noirlab.edu}
\affiliation{NSF NOIRLab, 950 N. Cherry Ave., Tucson, AZ 85719, USA}

\author[0000-0003-1704-0781]{Gregory Tarl\'{e}}
\email{gtarle@umich.edu}
\affiliation{University of Michigan, Ann Arbor, MI 48109, USA}

\author{Benjamin Alan Weaver}
\email{benjamin.weaver@noirlab.edu}
\affiliation{NSF NOIRLab, 950 N. Cherry Ave., Tucson, AZ 85719, USA}

\author[0000-0001-5381-4372]{Rongpu Zhou}
\email{rongpuzhou@lbl.gov}
\affiliation{Lawrence Berkeley National Laboratory, 1 Cyclotron Road, Berkeley, CA 94720, USA}

\author[0000-0002-6684-3997]{Hu Zou}
\email{zouhu@nao.cas.cn}
\affiliation{National Astronomical Observatories, Chinese Academy of Sciences, A20 Datun Rd., Chaoyang District, Beijing, 100012, P.R. China}

\begin{abstract}
We present details regarding a composite spectrum of quasar (QSO) absorption line systems. In this composite spectrum we identify more than 70 absorption lines, and observe oxygen and hydrogen emission features at a higher signal-to-noise ratio than any previous study. As the light from a distant quasar travels towards an observer, it may interact with the circumgalactic medium of an intervening galaxy, forming absorption lines. In order to maximize the signal of these absorption lines, we have selected a sample of 239{,}174 quasar spectra from the second data release of the Dark Energy Spectroscopic Instrument (DESI), each identified to have absorption lines resulting from such an interaction. By stacking these spectra in the restframe of the absorption, and calculating a median composite spectrum, we are able to isolate and enhance these absorption lines. We provide a full atlas of all detected absorption and emission lines as well as their fit centroids and equivalent width values. We additionally provide analysis of composites constructed using subsamples of absorbers in both redshift and \Mgii\ equivalent width bins. We find that there is minimal redshift evolution across the considered absorber samples. Additionally, we find that dense gas signatures, such as the presence of Na I absorption, are strongest in the highest equivalent width absorber sample, and that the strengths of the Si IV and C IV absorption lines show a relatively weak correlation with Mg II equivalent width. This atlas should aid in future studies investigating the compositions and physical conditions of these absorbers.
\end{abstract}

\section{Introduction}
Shortly following the discovery of quasars (or ``QSOs'') in the early 1960s \citep[e.g][]{QSODiscoveryI,QSODiscoveryII}, absorption lines were identified in their spectra at redshifts distinct from that of their emission. \citet{Wagoner} and \citet{BahcallSpitzer} were among the first to propose that these lines were caused by interactions with gas in the extended halos, i.e. the circumgalactic environments (CGM), of intervening galaxies that were within the line of sight to the more distant quasar. Studies of the CGM are summarized in \citet{CGM}.

Absorption lines in QSO spectra are commonly divided into two groups, parameterized by the velocity offset between the background QSO and foreground absorption environment, as determined using their respective redshifts:
\begin{equation}
    \rm{v}_{\rm off} = c \frac{z_{\rm QSO}-z_{\rm ABS}}{1+z_{\rm QSO}}
\end{equation}
Although the exact transition in velocity space between different types of absorbers is a complex matter, absorbers at \voff\ $ > 3{,}500$ \kms\ are generally considered to be intervening absorbers, whereas those systems at velocity offsets less then 3500 \kms\ are considered to be ``associated" with the QSO host galaxy \citep[e.g.][]{York2006,ZhuMenard2013,Khare2014,Chen2020,Ting2020,Abhijeet2021,Wu2025}.

The host galaxies of the circumgalactic environments that result in intervening absorption line systems are commonly too dim to be observed, particularly those at redshifts beyond $z=1$ \citep[e.g.][]{AbsHost1,AbsHost2,AbsHost3,AbsHost4}. As such, the produced absorption lines are one of very few methods by which we can gain insight into the gas content and other physical characteristics of these galactic environments.

The construction of median composite spectra is a technique frequently used in astronomy. By combining a large sample of similar objects and determining the median value at each wavelength value, one can reveal common features, and significantly increase signal-to-noise \citep[e.g.][]{QSOComp-SNR-Francis1991,QSOComp-Gen-HST}. In the world of large sky surveys, composite quasar spectra are frequently calculated in order to study emission or absorption line characteristics \citep[e.g.][]{QSOComp-Gen-FIRST, VandenBerk2001, VandenBerk2008,QSOComp-Gen-SDSSIV}, to characterize spectral energy distributions \citep[e.g.][]{QSOComp-SED-Sambruna1996,QSOComp-SED-Richards2006,QSOComp-SED-Mullaney2011}, and to determine the nature of different classes of quasars such as red quasars \citep[e.g.][]{QSOComp-Red-Selsing2016,Fawcett2023} or BAL quasars \citep[e.g.][]{QSOComp-BAL-Sprayberry1992,QSOComp-BAL-Gallagher2007}.

In this paper we calculate the composite spectrum of a large set of intervening absorption line systems. By shifting the QSO spectra that host these intervening absorbers into the restframe of their absorption, we are able to remove both the obscuring effect of the broad emission lines of the QSO, as well as reduce the innate noise in the individual data. Similar studies have been performed in order to search for high ionization state absorption lines \citep{AbsCompOVI}, or to characterize the physical conditions and metal enrichment of the intergalactic medium \citep{York2006,AbsCompLyA}, or to explore the physical properties of intervening \Mgii\ gas clouds \citep[e.g][]{ZhuMenard2013,Khare2014,Ting2017,Chen2020,Abhijeet2021}.

A number of studies have utilized absorber composite spectra in order to further investigate the relationship between \Mgii\ absorbers and star formation, in both the cases of associated and intervening absorption systems. \citet{Wild2007-CaIIComposites} constructed composites using both Ca II and Mg II absorbers in order to detect the [O II] 3727, 3729$\textrm{\AA}$ emission lines. \citet{Menard2011} then proposed, and characterized, a relationship between the rest-frame equivalent width of Mg II absorbers and the star formation rate, as measured using the [O II] line. This relationship has been further investigated by \citet{ShenMenard-OII-StarForm} and \citet{Khare2014} which both use samples of associated absorption systems, and find increased [O II] luminosity as being associated with the presence of an associated absorption line system.

This result was challenged by \citet{Lopez2012-OII} which argued that the claimed relationship was likely due to fiber loss of [O II] flux from galaxies at larger impact parameters, which would also naturally produce weaker \Mgii\ absorbers. This claim is furthered by \citet{Ravi2018-OIIEmission} which finds that for a given fiber size the luminosity of the [O II] line strongly depends on the equivalent width of Fe II 2600$\textrm{\AA}$ absorption, and that the strong correlation of [O II] luminosity with \Mgii\ equivalent width may be mainly driven by systems with large Fe II equivalent widths.

A similar analysis of galaxies in the foreground of QSO spectra was carried out in the Sloan Digital Sky Survey \citep[e.g.][]{York2012-Halpha,Straka2013-EmissionLines,Straka2015-EmissionLines}. By detecting sets of QSO-galaxy pairs observed in the same fiber these studies were able to measure the emission features directly in individual spectra without the need for composite spectra.

Absorber composite studies commonly utilize a catalog of absorbers formed based on the detection of a single high-signal absorption species. The choice of this high-signal species is significant as it, along with the wavelength coverage of the spectra utilized, determine the wavelength coverage of the computed composite spectra. In this study we choose to utilize an input catalog of Mg II absorbers \citep{MgIICat} detected in spectra collected by the Dark Energy Spectroscopic Instrument (DESI) \citep{DESI_Instrument,DESI-Instrumentation-2022,DESI-SpectroPipeline-2023,DESI-Corrector-2024,DESI-Fibers-2024}. Details regarding the daily operations of the DESI survey are presented in \citet{SVOPS}.

DESI has recently completed the processing of its second data release, containing spectra from the first three years of the survey \citep{DESI2024.I.DR1,DESI-DR2BAO-2025}. The QSO catalog produced using this data release has a total of  2{,}456{,}178 entries, and the details of their selection and cataloging are given in \citet{QSOTS} and \citet{QSOVI}. Note that we restrict our parent QSO sample to only those spectra that were originally targeted as QSOs in the interest of sample purity. The Mg II absorption system catalog constructed from this parent QSO sample has a total of 539{,}151 entries, detected in 392{,}449 total spectra, and its construction is detailed in \citet{MgIICat} and \citet{MgIIEBV}.

In this paper we will utilize this sample in order to construct an extremely high signal-to-noise QSO absorption line system composite. From this composite, we will create an atlas of detected emission and absorption species and measure their centroids relative to catalog values, as well as their equivalent width values. In doing so, we will provide a more complete understanding of these systems than has previously been possible. We will additionally analyze the composite spectra constructed using subsamples of absorbers as selected by their \EWFL\ and z$_{\textrm{\tiny ABS}}$ value.

This paper will be organized in the following fashion: In \S2 we will detail the construction of the absorber composite including our sample selection and the composite construction methodology. In \S3 we will analyze the full sample composite, measuring all detected lines with a particular focus on the emission features we find. We will further consider subsample composites investigating the relationship of W$_0$ with respect to \EWFL\ and z$_{\textrm{\tiny ABS}}$. In \S4 we will discuss challenging line identifications, as well as verifying the nature of the emission features detected and considering their significance. In \S5 we present our conclusions.

\section{Data Methods}
In this section we will outline the selection of our sample of intervening absorption systems, as well as that of our control sample of QSOs with no absorption systems. We will then detail the methods by which we construct our median composite spectrum, and the methods by which we identify and fit the absorption lines present in the composite spectrum.

\subsection{Sample Selection}
In order to construct our composite spectrum, we must first select a suitable sample of absorbers from the full population of 539{,}151 absorbers initially detected in the DESI DR2 data. Firstly, we remove all associated absorption systems, that is absorbers with velocity offset values less than 3500 \kms, as the restframe of these absorbers is similar to the restframe of their background QSOs, and the QSO emission lines present could influence the resulting composite spectrum. 

Next, we make the choice to incorporate only QSO spectra in which a single \Mgii\ absorber was detected. This ensures a smoother continuum in the resulting composite, as the effect of absorption lines from absorption systems at other redshifts, as well as the effect of the emission lines of each background QSO, should be minimized or mitigated entirely. Both of these features could be amplified if we were to use QSO spectra in which multiple absorbers were detected, particularly at the edges of our redshift coverage where fewer spectra are available. These choices reduce our sample to 239{,}174 single intervening absorber QSO spectra.

Notably, we do not expect this latter choice, to use only QSO spectra in which a single \Mgii\ absorber was detected, to affect the science results of this study, as the natures of all intervening absorption systems are similar, and the presence of multiple absorption systems in a single QSO spectrum is solely the result of the line of sight to the QSO. As such, this choice is made primarily to simplify the reduction of the composite spectrum, which we will detail in \S2.3. 

\begin{figure}[ht!]
\plotone{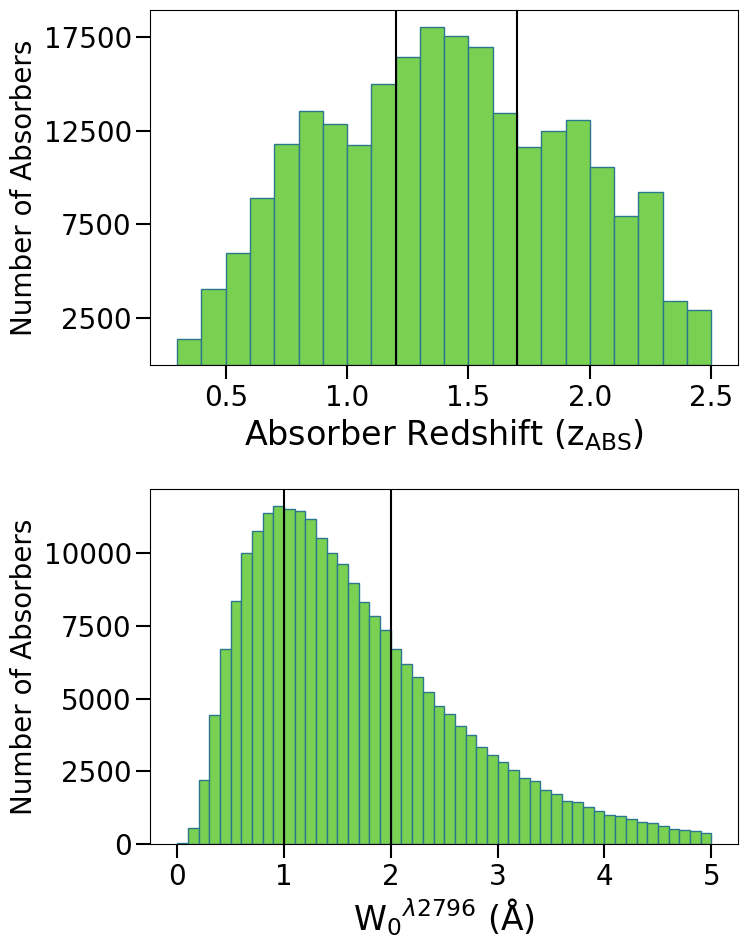}
\caption{Histogram of absorber redshift values for the 239{,}174 absorption line systems in our sample. Bins are 0.1 wide in redshift space. The overlaid black vertical lines indicate the subsamples selected in sections \S 3.2 and 3.3.}
\label{ZABS-Hist}
\end{figure}

As the absorption systems present in these spectra have been identified using the \Mgii\ absorption doublet at 2796$\textrm{\AA}$ 2803$\textrm{\AA}$, and DESI spectra have an observed wavelength coverage of [3600$\textrm{\AA}$, 9824$\textrm{\AA}$], the range of possible absorption systems redshifts is z$_{\textrm{\tiny ABS}} \approx$ [0.3, 2.5]. Figure \ref{ZABS-Hist} presents the redshift distribution for these absorbers. Fewer absorbers are detected at the low and high ends of this distribution due in large part to the underlying QSO redshift distribution in DESI, as well as increased noise at the edges of the DESI spectrum \cite[e.g.][]{QSOTS}. Additionally, there are slight dips in the distribution around z $\approx$ 1.1 and z $\approx$ 1.8, which we interpret to be due to the DESI spectrograph crossover regions \citep{DESI_Instrument,MgIICat}.

Figure \ref{ZABS-Hist} also presents the distribution of \Mgii\ rest-frame equivalent width (\EWFL) values present in our sample. The median \EWFL\ value in our sample is $1.487\textrm{\AA}$, and the 16th and 84th percentile values are $0.755\textrm{\AA}$ and $2.723\textrm{\AA}$ respectively. We choose not to make any selection on the rest-frame equivalent width in the interest of maximizing sample size.

Next, we construct a control sample of DESI QSOs without any detected \Mgii\ absorbers. The total population of such objects in DESI DR2 is 2{,}061{,}617 spectra. From this sample we first remove the 25 percent of QSOs that have lowest g-band flux. This is done as the completeness of our \Mgii\ detection approach is reduced for these fainter spectra \citep{MgIICat}, and as such we can be less confident that these spectra do not have undetected \Mgii\ absorbers. Additionally, this cut results in the distributions of g-band flux values between QSO spectra with, and without, \Mgii\ absorbers being in greater agreement as the population of spectra with detected absorbers tends to be brighter than that of spectra with no detected absorbers.

For each single-absorber QSO spectrum in our sample, we then select a matching control QSO spectrum. This selection is performed for each absorbed QSO spectrum by determining which DESI QSO with no detected \Mgii\ absorbers minimizes the total difference in redshift and {\tt TSNR2\_QSO} value. {\tt TSNR2\_QSO} is a measurement of the signal-to-noise ratio of the spectrum, where certain regions are more heavily weighted according to the expected spectral characteristics of the target class \citep{PipelinePaper}. Each control QSO spectrum is matched exclusively to a single-absorber QSO spectrum, i.e. each member of the control sample is unique. Comparison of this control sample will enable confirmation of the source of any features in the single-absorber composite spectrum, as well as aiding in the fitting of these features. 

\begin{figure}[ht!]
\epsscale{1.15}
\plotone{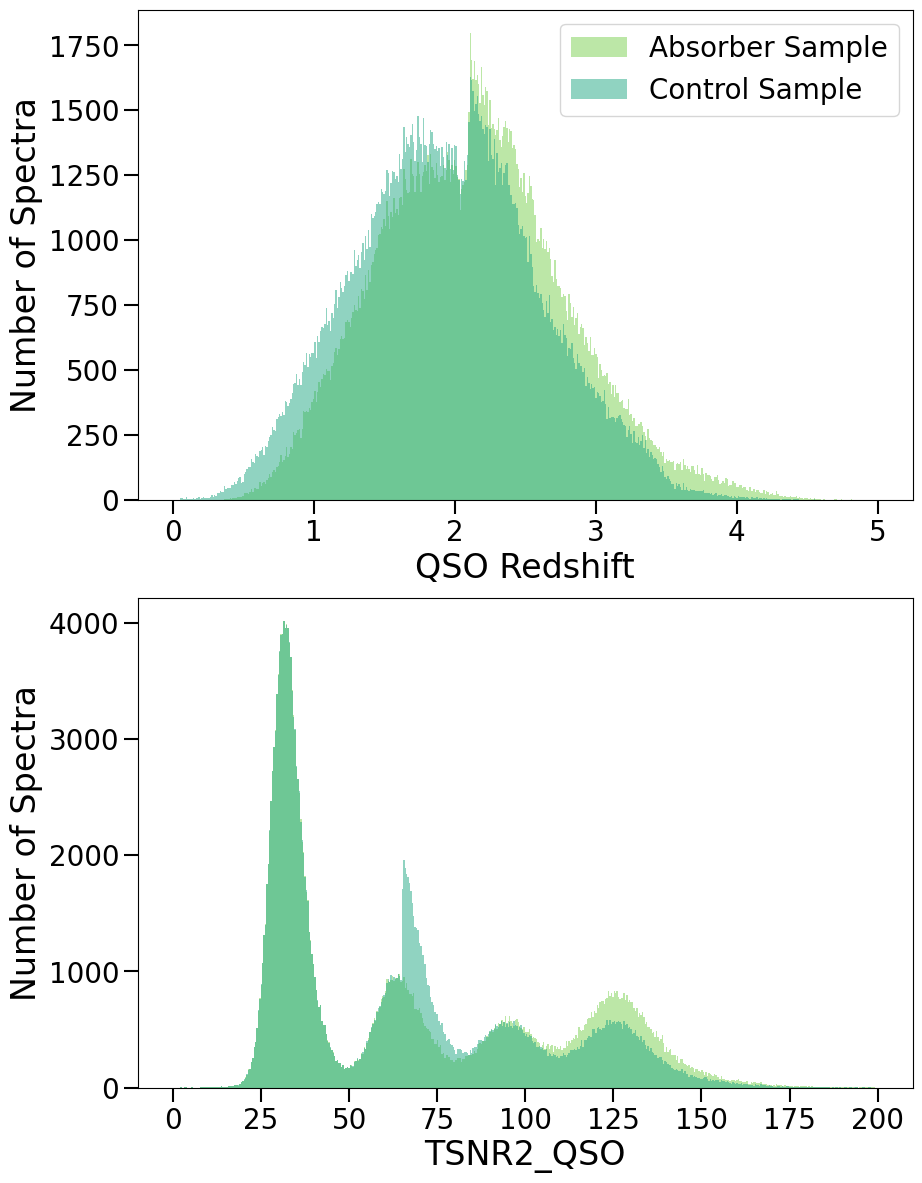}
\caption{Histograms displaying the distributions of the redshift (top), TSNR2\_QSO (middle) and W2 band flux values for both our absorber (light green) and control (blue) sample. Each histogram is drawn with 500 equally sized bins.}
\label{ControlCompare}
\end{figure}

Figure \ref{ControlCompare} displays the distributions of multiple parameters of our absorber and control samples. Comparing them we can note that the absorber sample tends to have slightly higher values of QSO redshift and TSNR2\_QSO. This is due to a bias in detecting Mg II absorbers, as QSOs above redshift z = 2.1 are targeted to receive additional observations in DESI \citep{TargetingPaper}. Multiple observations of the same object then increase signal-to-noise ratio, and make it easier to detect an absorption line system \citep{MgIICat}. Noting this, we consider our two samples to be in strong enough agreement to enable the analyses we perform in this paper.

\subsection{Composite Construction}
\label{sec:compositeconstruction}

\begin{figure}[ht!]
\plotone{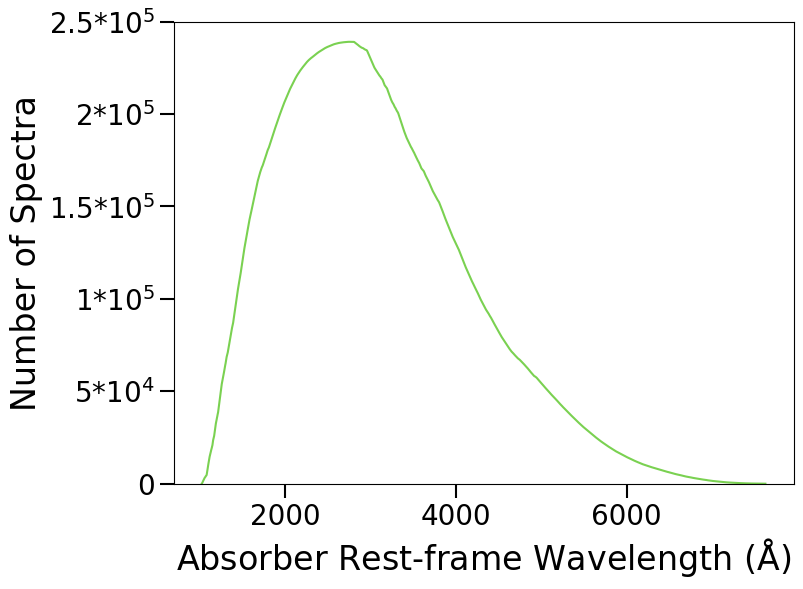}
\caption{The number of spectra, that when resampled into the absorber restframe, cover each pixel of the composite's wavelength grid. The distribution peaks in the range [2795$\textrm{\AA}$, 2804$\textrm{\AA}$], which is the location of the \Mgii\ doublet, and is available in the restframe of all 239{,}174 absorption systems.}
\label{WaveCov}
\end{figure}

Having determined our sample of absorbed QSO spectra, as well as our control sample, we can now consider the methods by which we calculate the composite spectrum for each. As previously mentioned, the wavelength coverage of DESI spectra is [3600$\textrm{\AA}$, 9824$\textrm{\AA}$] and the approximate redshift range of our absorption line systems is z$_{\textrm{\tiny ABS}} \approx$ [0.3, 2.5]. As the rest-frame wavelength coverage of a z$_{\textrm{\tiny ABS}}$ = 0.3 absorber is $\approx$ [2770, 7557], and that of a z$_{\textrm{\tiny ABS}}$ = 2.5 absorber is $\approx$ [1028$\textrm{\AA}$,2806$\textrm{\AA}$], this informs a natural wavelength range of [1030$\textrm{\AA}$, 7500$\textrm{\AA}$] for our absorber restframe composite. Relatedly, we choose a pixel spacing of of 0.4$\textrm{\AA}$, corresponding to the native DESI wavelength spacing of 0.8$\textrm{\AA}$, shifted into the restframe of a  z$_{\textrm{\tiny ABS}}$ = 1.0 absorption system, a value close to the center of our redshift distribution. Figure \ref{WaveCov} shows the number of spectra that when shifted into the absorption restframe contribute to each pixel of the composite spectrum.

\begin{figure*}[ht!]
\epsscale{1.15}
\plotone{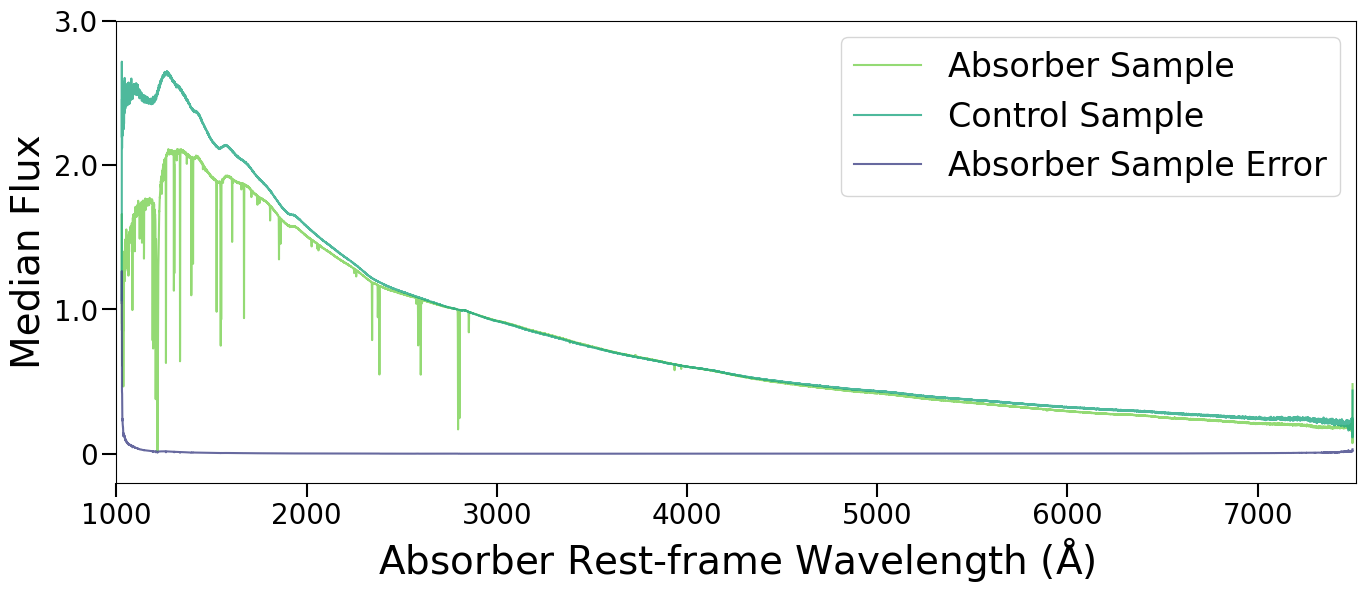}
\caption{Median composite spectra calculated for both our sample of QSO spectra with a single detected \Mgii\ absorber, and our matched control sample of QSO spectra with no detected absorbers. Also presented is the error spectrum associated with the absorber sample composite. Note that we do not show the error spectrum associated with the control sample, as the two error spectra would appear nearly identical.}
\label{Initial-Comp}
\end{figure*}

Next, we shift each spectrum in our sample into the restframe of its detected absorption system by dividing its wavelength coverage by (1+ z$_{\textrm{\tiny ABS}}$). We then resample the flux values of the spectrum onto the common rest-frame wavelength grid using the Python package {\tt scipy.signal}, which utilizes FFT transformations to perform the resampling \citep{Scipy}. In order to normalize our individual spectra, we divide by the median flux in two regions immediately adjacent to the \Mgii\ absorption doublet, [2760$\textrm{\AA}$, 2780$\textrm{\AA}$] and [2820$\textrm{\AA}$, 2840$\textrm{\AA}$]. Both of these regions are available in the restframe for the majority of our sample, however at the edges of our redshift distribution only one may be present.

We perform this process of shifting into the restframe, resampling flux values, and normalizing via the regions adjacent to the \Mgii\ absorption doublet for all 239{,}174 entries in our sample of QSO spectra that host one detected absorption line system. We then repeat this process for our control sample, using the absorption system redshift of the absorbed spectrum to which each control group member was matched. Finally, we calculate the median value in each pixel of our common rest-frame wavelength grid, producing our composite spectrum. We determine the error in our composite spectrum by taking the difference between the 16th and 84th percentile value in each pixel and dividing by the square root of the number of spectra which, when shifted into the restframe of the absorption system, cover that pixel, as shown in Figure \ref{WaveCov}.

In Figure \ref{Initial-Comp}, we present the results of this procedure for both our sample of spectra with a single detected absorption system, and our control sample. We can immediately note that the control sample composite is significantly brighter than the absorber sample composite at wavelengths less than 2000$\textrm{\AA}$, and slightly brighter at wavelengths greater than 5000$\textrm{\AA}$. A number of absorption lines are immediately evident in the absorber sample composite, including the \Mgii\ doublet at $\sim$ 2800$\textrm{\AA}$, C IV doublet at $\sim$ 1550$\textrm{\AA}$ and a broad \LyA\ absorption line at $\sim$ 1215$\textrm{\AA}$. In the following subsection, we will detail the methods by which we identify and fit these lines, and the many additional lines that are present. We note here that the composite spectrum data plotted in Figure \ref{Initial-Comp}, and the data of all other plotted composite spectra, is publicly available as part of the zenodo entry  associated with this paper \footnote{https://doi.org/10.5281/zenodo.17497238}.

\subsection{Composite Reduction \& Line Fitting}
Having calculated our composite spectrum, we can now consider the process of line identification. Rather than automating this step, we choose to perform a precise visual inspection, wherein the absorber sample composite was carefully scanned by eye in order to identify possible lines. Observed lines were then identified, and confirmed, using resources such as atomic spectra atlases, like those provided by \citet{Morton91, Morton03} and \citet{NIST}, as well as lists of absorption lines from similar projects and references therein \citep[e.g.][]{AbsCompOVI,VandenBerk2001,AbsCompLyA}. Note that hereafter we will refer to \citet{Morton03} as M03 and \citet{NIST} simply as NIST.

In total we observe 72 unique absorption lines, arising from 26 unique absorption species. Additionally, we observe 7 {\em emission} lines corresponding to the transitions: [O II], H$\gamma$, H$\beta$, [O III] and H$\alpha$.

In order to cleanly fit these atomic lines, we will need to normalize our composite spectrum, for which we can utilize the control sample composite detailed in \S\ref{sec:compositeconstruction}. As previously noted, the overall shape, ignoring absorption species, of the control and ``absorber'' composite spectra are quite similar, however the control sample composite is significantly brighter at wavelengths less than 2000$\textrm{\AA}$. \textbf {This difference in brightness is due to reddening from the dust associated with the absorbing gas, details regarding the magnitude of this absorption can be found in \citet{MgIIEBV}}. 

\begin{figure*}[ht!]
\plotone{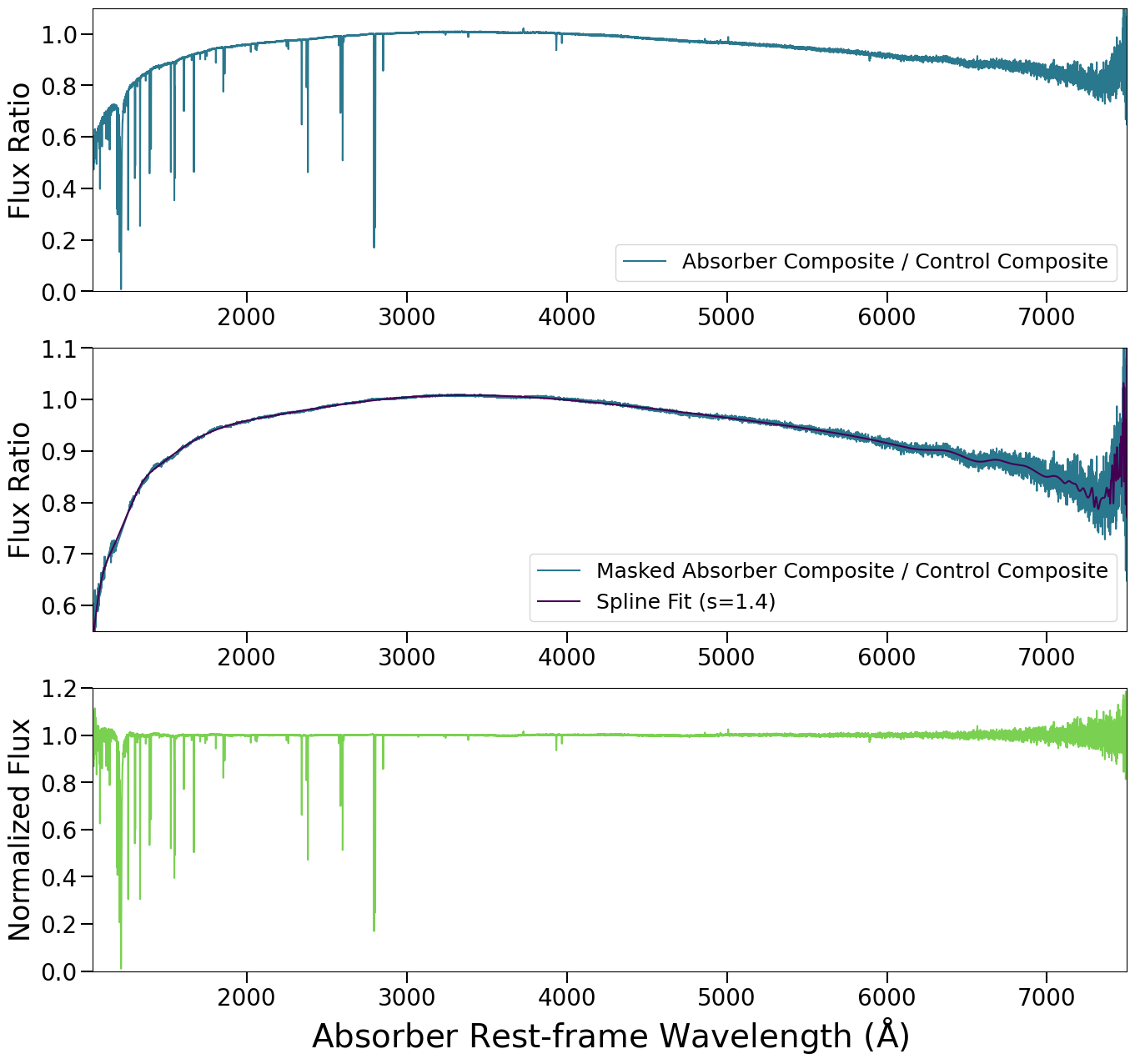}
\caption{Visualization of the absorber sample continuum reduction method. \textit{Top:} The result of dividing the absorber sample composite by the control sample composite, as seen in Figure \ref{Initial-Comp}.  \textit{Middle:} The result of masking the regions around the detected absorption and emission lines as described in the text. The purple line then shows the result of fitting a cubic B-spline with smoothness = 1.4 to the masked data. \textit{Bottom:} The continuum-reduced absorber composite, calculated by dividing the product of the spline fit from the middle panel and the initial control sample composite.}
\label{SplineFit}
\end{figure*}

To better fit the control sample composite to the absorber composite, we first divide the absorber sample composite by the control sample composite, resulting in the spectrum seen in the top panel of Figure \ref{SplineFit}. We then mask all pixels within 10$\textrm{\AA}$ of all detected emission and absorption lines, increasing this to 40$\textrm{\AA}$ for the broader \LyA\ line. 

Next, we fit a cubic B-spline, as implemented in the Python package {\tt SciPy}, to the masked composite spectrum. The resulting fit can be seen in the middle panel of Figure \ref{SplineFit} \citep{Scipy}. In order to avoid overfitting or underfitting the masked composite, we must carefully choose a smoothness value for our spline. We do so by testing a range of possible values, and calculating the sum of squared differences from a $y = 1$ line in the previously masked regions around our detected emission and absorption lines. We find that this quantity steadily decreases with increased smoothing, up to a value of 1.4, at which point further smoothing has little effect in the line regions, and instead begins to underfit the broad features of the composite continuum. For this reason, we select a smoothing value for our spline of $s = 1.4$.

\begin{figure*}[ht!]
\epsscale{1.15}
\plotone{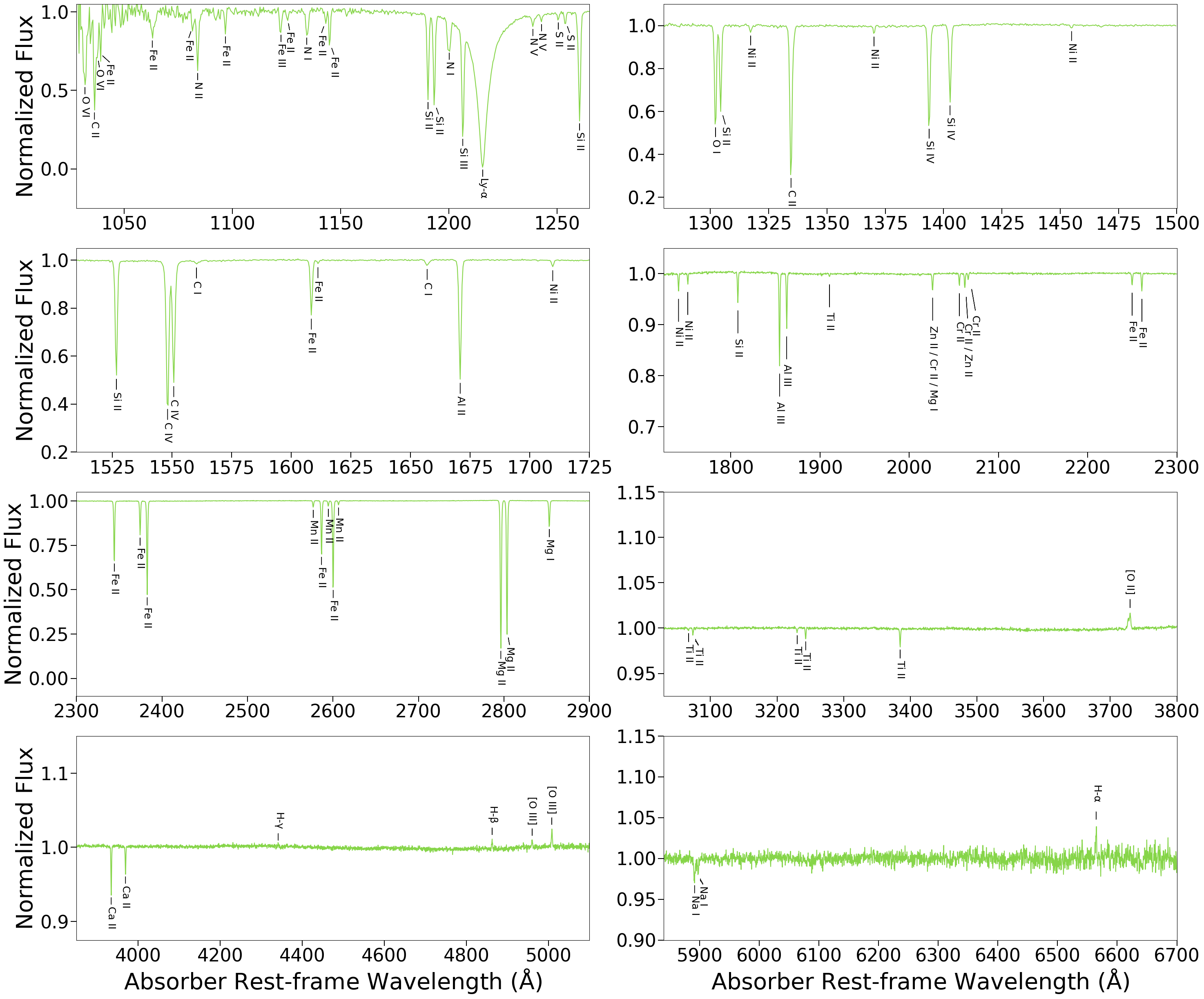}
\caption{Labeled and horizontally stretched version of the reduced absorber sample median composite spectrum, as plotted in Figure \ref{SplineFit}. Detected absorption and emission lines are labeled, and lines that are blends of multiple species have each species listed. Note that the scale on both axes varies.}
\label{LineCompReduced}
\end{figure*}

By then dividing the absorber sample composite by the product of our spline fit and the control sample composite, we successfully reduce our absorber sample composite, such that it is suitable for line fitting. This reduced composite spectrum can be seen in the bottom panel of Figure \ref{SplineFit} and a zoomed in and labeled version of this spectrum is shown in Figure \ref{LineCompReduced}.

\begin{figure*}[ht!]
\epsscale{1.15}
\plotone{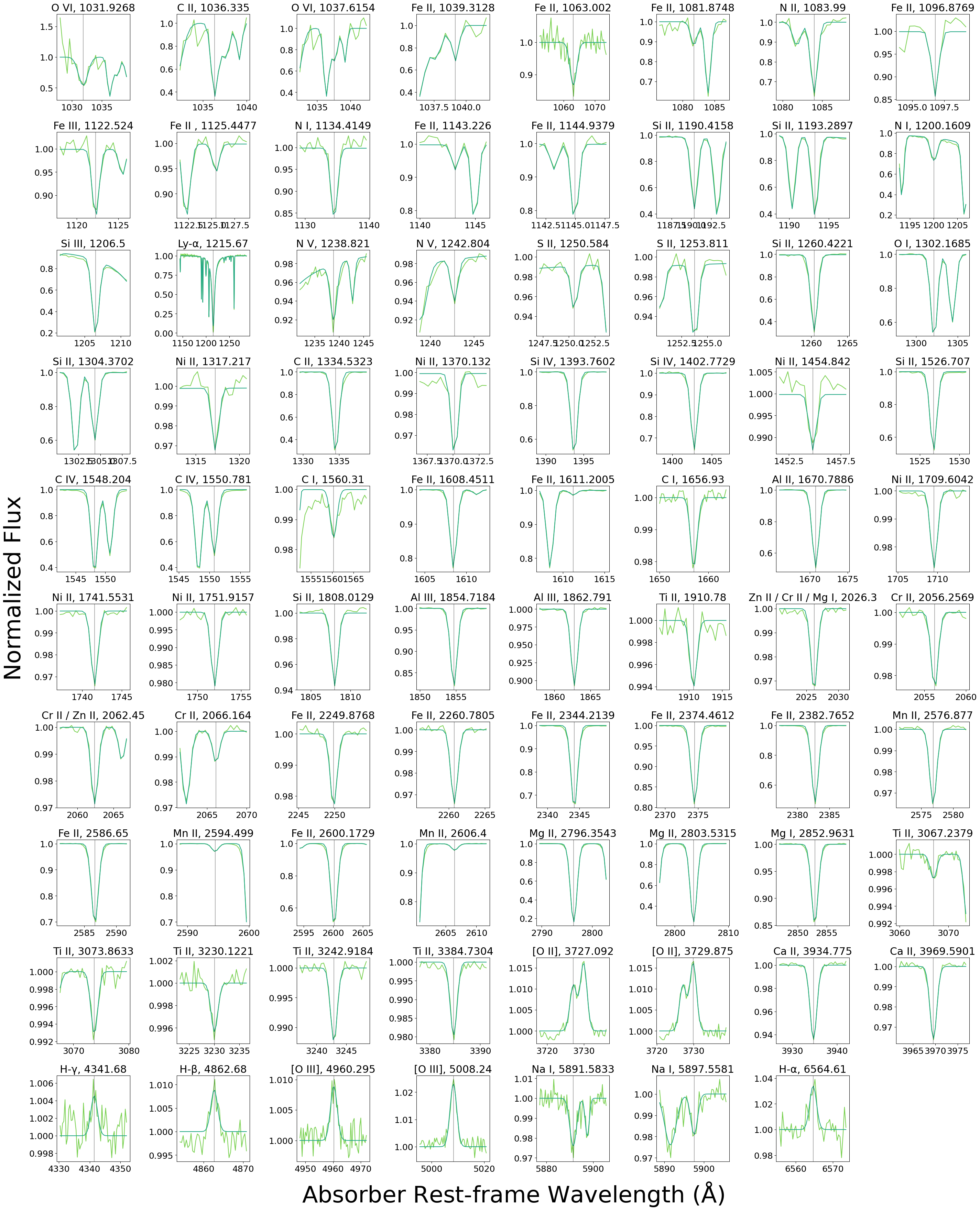}
\caption{Grid view of the output of our line fitting procedure. A subplot is drawn for each noted absorption or emission line. The absorber composite is shown in light green, and the model fit shown in blue. Each line is plotted in a 10$\sigma$ range around its centroid value. Subplots are titled by the line's vacuum wavelength, as given in Tables 1 \& 2.}
\label{LineFitGrid}
\end{figure*}

Having now reduced our absorber composite spectrum, we can consider the matter of line fitting. We have designed our own line-fitting procedure, in which the majority of lines are fit using a single {\tt Gaussian1D} model, as implemented in {\tt Astropy}, which is initialized at the catalog wavelength of the atomic line \citep{Astropy}. For lines whose catalog wavelength values are within 4.0$\textrm{\AA}$ of a second line, such as the C IV and [O II] doublets, or O I 1302$\textrm{\AA}$ and Si II 1304$\textrm{\AA}$, we instead use a compound model composed of two {\tt Gaussian1D} models. The lines in the region around \LyA\ require a special case solution: first, due to its width, we fit the \LyA\ line itself using a {\tt Voight1D} profile, and then form a compound model that includes all other lines in the region from 1200 to 1255$\textrm{\AA}$. A visualization of the output of our line fit procedure for all detected lines is shown in Figure \ref{LineFitGrid}.

After fitting each line, we record the parameters of the fit, as well as calculating the equivalent width of the line. In order to determine the uncertainties in these values, we draw 10{,}000 resampled composite spectra according to the error spectrum associated with our reduced absorber composite and perform our fit procedure, again recording the line parameters and equivalent width. In the next section we will detail the results of this process and discuss our full atlas of observed absorption and emission lines.

\section{Results and Discussion}

In this section we first discuss the structure of the output tables we have constructed using our line fitting results, as well as the nature of the absorbers that could lead to such a wide range of ionization potentials. We next consider the results of two subsample analyses, where we have used both \EWFL and $\textrm{z}_{\textrm{\tiny ABS}}$ to split the full sample into three separate bins.

\subsection{Line Fitting Results}
The fit line centroid values and $W_0$ values for all fit line species are shown alongside their catalog vacuum wavelength and oscillator strength values in Tables~\ref{table:abslineatlas} and ~\ref{table:emlineatlas}. All vacuum wavelengths and oscillator strength values are sourced from NIST when available. Section \S 3.2 details which lines are instead sourced from M03 due to data availability. We can immediately note that the uncertainties in $W_0$ are highest at the edges of the wavelength coverage of the composite, and steadily decrease approaching the \Mgii\ lines, which have extremely precise measurements. Examining the $W_0$ values, we note that the \LyA\ line is by far the strongest, and other strong lines include the Mg II doublet (2796$\textrm{\AA}$, 2803$\textrm{\AA}$), Si III 1206$\textrm{\AA}$, the C IV doublet (1548$\textrm{\AA}$, 1550$\textrm{\AA}$), and O VI 1032$\textrm{\AA}$, although this line has large uncertainties. The centroids of all fit lines are within 1$\textrm{\AA}$ of the reference value, and many are within 0.1$\textrm{\AA}$ or less, suggesting that our fits are reliable and species identifications accurate.

We note that our list contains a number of blended line systems, these include both multiplet blends such as N I 1134$\textrm{\AA}$, N I 1200$\textrm{\AA}$ and Ti II 1910$\textrm{\AA}$ as well as inter-species blends such as lines we observe at 2026$\textrm{\AA}$ and 2062$\textrm{\AA}$, which are composed of Zn II, Cr II, Mg I and Cr II and Zn II respectively. For each of these blends, the table wavelength we provide is an un-weighted average of the components. However we do provide the oscillator strength values for each component line as well as the individual vacuum wavelength values in a table footnote.

Examining the complete list of fit lines, we note that they cover a large range of ionization potentials, and as such it is impossible that they could arise from a single gas environment. Their fit line centroids being in such strong agreement is then a result of DESI's velocity resolution, which is $\sim150 $ \kms\  at $3600\textrm{\AA}$ and $\sim50 $ \kms\ at $9800\textrm{\AA}$, as well as possible line of sight projection effects and the peculiar velocities of different CGM gas environments. This result was recently demonstrated in simulations by the FOGGIE project \citep{Peeples2019} and seen experimentally by comparing \Mgii\ and O VI correlation functions \citep{Nielsen2017}.

\subsection{\EWFL\ Subsample Composite Spectra}

In addition to the analysis we have provided regarding the full sample composite, we felt it prudent to also consider the variance of the composite spectrum if it were instead constructed using only a subsample of the population. Naturally, these subsamples could be selected using potentially intriguing variables. In this section we consider qualitative results from this analysis as pertain to subsets selected using the Mg II absorption line strength, (\EWFL), and in the following section \S 3.3 we will consider subsets selected using the redshift of the systems.

In selecting a reasonable number of bins for our \EWFL\ subsamples we have decided to use three bins, which naturally maps to their containing ``weak", ``average" and ``strong" absorbers relative to our samples population. If dividing our sample evenly the 33rd and 66th percentile values of \EWFL\ are 1.111 and 1.932 respectively. In the interest of having more cleanly defined bin edges we have selected three bins: \EWFL\ $\leq 1.0 \textrm{\AA}$, $ 1.0 \textrm{\AA} < $ \EWFL\ $\leq 2.0 \textrm{\AA}$ and \EWFL\ $ > 2.0 \textrm{\AA}$. These bins contain approximately 27.6\%, 40.5\% and 31.9\% of the sample respectively.

\begin{figure*}[ht!]
\epsscale{1.15}
\plotone{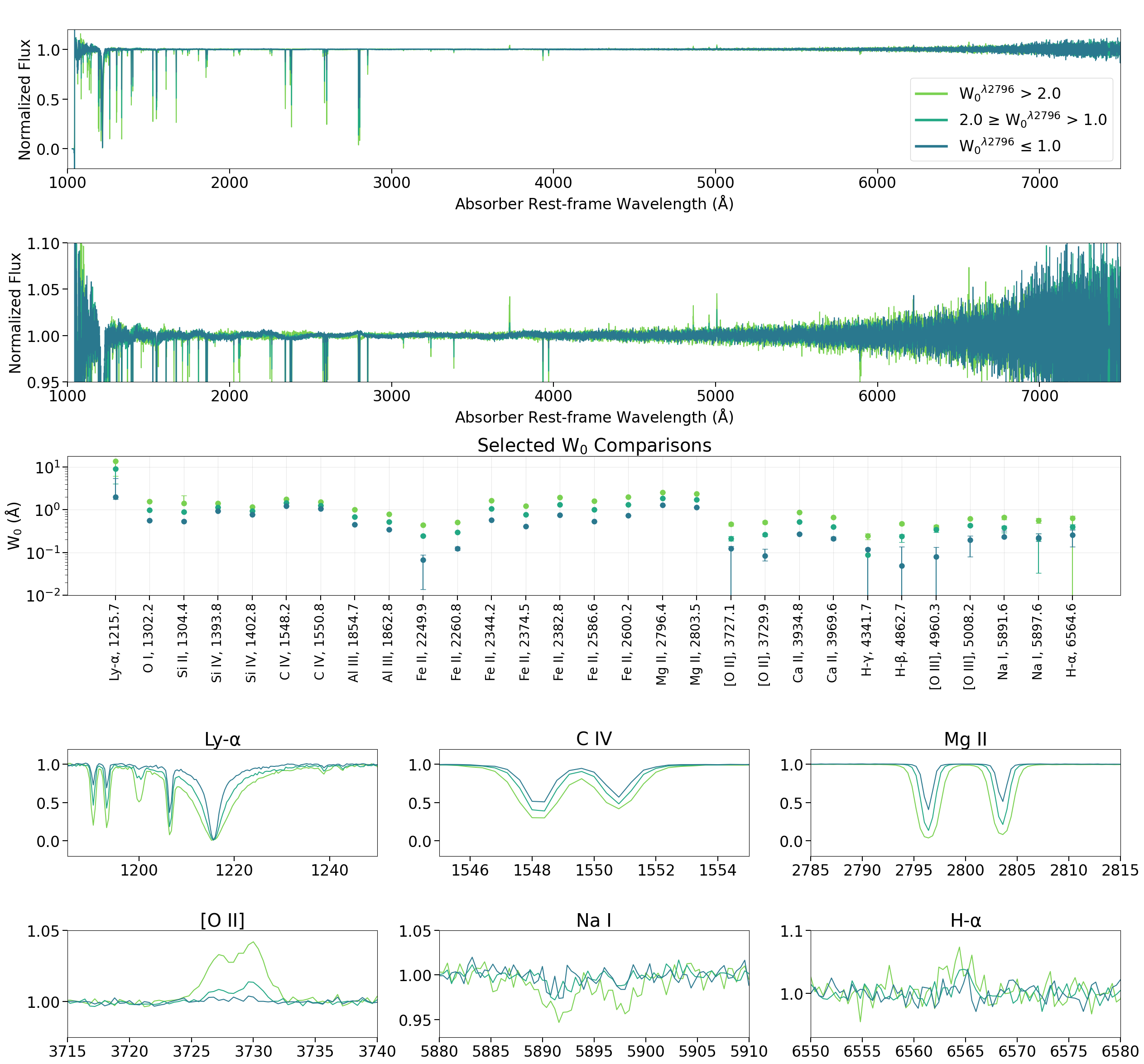}
\caption{Comparisons of \EWFL\ subsample composite spectra. \textit{Top}: Full view of subsample composite spectra, with the 3 selected bins indicated and overplotted. \textit{Second panel}: Zoomed in view of composite spectra, revealing weaker absorption species and emission features. \textit{Third panel}: Line strength measurement (W$_0$) comparisons for a group of selected line species. Note that the colors used correspond to the legend of the top panel. \textit{Bottom panels}: Cut-out views of specific regions and line species. Note again the colors correspond to the legend given in the top panel.}
\label{EW-bins}
\end{figure*}

Having selected our subsamples we can now repeat the processes described in \S2 in order to construct their reduced composite spectra. The results of doing so are shown in the top two panels of Figure \ref{EW-bins}. We can observe that all three composites cover approximately the same wavelength range as the full sample composite, indicating that absorbers in all three subsample bins can be found at essentially all absorber redshift values.

Examining the selected W$_0$ comparisons shown in the third panel of Figure \ref{EW-bins} we can note a few intriguing trends. Firstly, the W$_0$ values for Mg II 2796 found in this analysis are slightly greater than the value that would be suggested by their bin. We attribute this result to the significantly higher degree of noise that would be present in the individual spectra in which these absorbers were first detected. This result does suggest that the Mg II line detection procedure detailed in \citet{MgIICat} may, on average, result in an underestimation of \EWFL\ values.

Further investigating we can note that in all selected cases the strongest absorption lines are found in the highest \EWFL\ bin, this is also generally true for the emission species. However, looking at the Si IV and C IV lines we can observe that their W$_0$ values are not as strongly positively correlated with \EWFL\ as are the W$_0$ values of other species, such as Fe II or Ca II. That the strength of these lines does not depend as strongly on the \EWFL\ value reinforces the idea that they may trace different gas phases.

Examining the cut-outs shown at the bottom of Figure \ref{EW-bins} we can clearly see the effect selecting on \EWFL\ has on other line species. The \LyA\ line does not noticeably reduce in amplitude, as it is fully saturated, but its wings are significantly reduced in the lower \EWFL\ bins. Additionally, we can see that the N I 1200 line is strongly affected and nearly absent from the \EWFL\ $\leq 1.0 \textrm{\AA}$ bin.

By comparison, the C IV line, as previously noted, shows a relatively small reduction in line strength. Examining the emission lines we can see that [O II] strength is strongly affected by reduced \EWFL\ whereas H-$\alpha$ is not as strongly affected.

Finally, we can consider the Na I line, which is relatively unique in our sample, having an extremely low ionization potential (5.1 eV) and being the only absorption line we detect above 4000 $\textrm{\AA}$. The existence of Na I suggests a particularly dense gas environment, unlike that in which we would expect to find Mg II or C IV absorption. Examining both the W$_0$ values and the cut-out we can see that Na I is strongest in the highest \EWFL\ bin and is quite weak in both the middle and low bin. This suggests that the lines of sight that intercept the dense gas environments that could be host to Na I also tend to intercept multiple strong Mg II gas clouds.

\subsection{Absorber Redshift Subsample Composite Spectra}

We can now consider separating our sample into bins based on the absorber redshift (z$_{\textrm{\tiny ABS}}$). Mirroring our decision in the previous section we choose to use 3 distinct bins. The exact 33rd and 66th percentile z$_{\textrm{\tiny ABS}}$ values in our sample are 1.16 and 1.63 respectively. We choose to use bins split at z$_{\textrm{\tiny ABS}}$ = 1.2 and 1.7. When considering redshift bins we inherit absolute upper and lower limits from the wavelength coverage of DESI spectra, these are z$_{\textrm{\tiny ABS}}$ = 2.505 and z$_{\textrm{\tiny ABS}}$ = 0.287 respectively. So our three redshift bins are $2.505 > \textrm{z}_{\textrm{\tiny ABS}} > 1.7$, $1.7 \geq \textrm{z}_{\textrm{\tiny ABS}} > 1.2$ and $1.2 \geq \textrm{z}_{\textrm{\tiny ABS}} > 0.287$. These bins contain approximately 29.8\%, 34.5\% and 35.7\% of the sample respectively.

In comparison to our EW subsample analysis, where we saw that the wavelength coverage of the resulting composite spectra was similar to the full sample composite spectrum, selecting on z$_{\textrm{\tiny ABS}}$ directly affects the wavelength coverage of the resulting composite. As a result the three composite spectra we will construct using our z$_{\textrm{\tiny ABS}}$ subsamples will only share a common wavelength region from 1636 $\textrm{\AA}$ to 3638 $\textrm{\AA}$, restricting our analysis to only those lines that fall within this region.

\begin{figure*}[ht!]
\epsscale{1.15}
\plotone{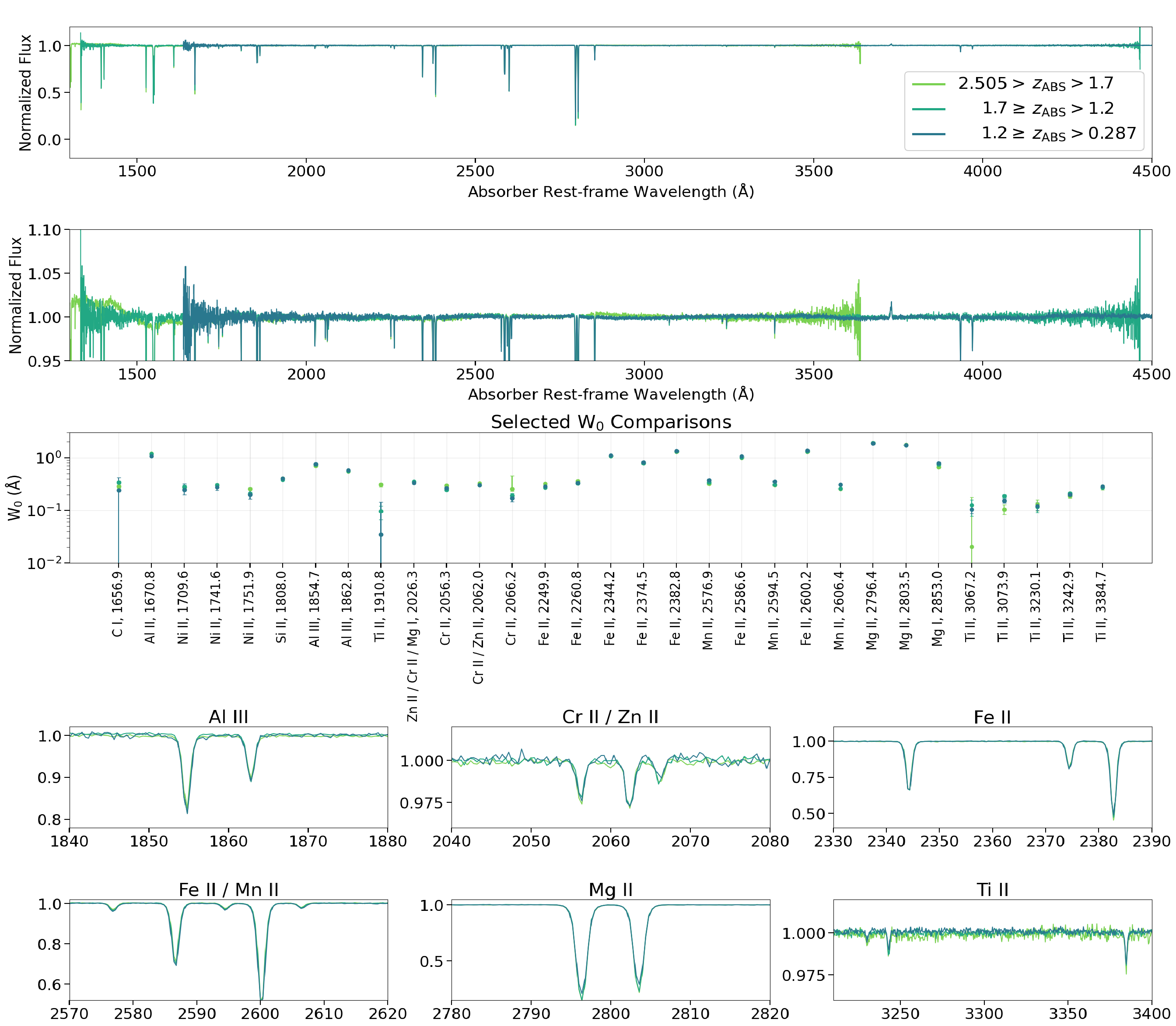}
\caption{Comparisons of z$_{\textrm{\tiny ABS}}$ subsample composite spectra. \textit{Top}: View of subsample composite spectra, with the 3 selected bins indicated and overplotted. Note that only the regions in which at least two of the composite spectra are available are plotted. \textit{Second panel}: Zoomed in view of composite spectra, revealing weaker absorption species and emission features. \textit{Third panel}: Line strength measurement (W$_0$) comparisons for a group of selected line species. Note that the colors used correspond to the legend of the top panel. \textit{Bottom panels}: Cut-out views of specific regions and line species. Note again the colors correspond to the legend given in the top panel.}
\label{z-bins}
\end{figure*}

The reduced composite spectra constructed using our z$_{\textrm{\tiny ABS}}$ subsamples are presented in Figure \ref{z-bins}. In the top two panels we have plotted the region in which at least two of the composites are available. This makes clear that the low redshift subsample composite only becomes available at 1636 $\textrm{\AA}$ and the high redshift subsample composite becomes unavailable at 3638 $\textrm{\AA}$.

Despite the limited wavelength coverage, we are still able to investigate the trends in W$_0$ for 31 absorption features, all of which are shown in the third panel of Figure \ref{z-bins}. We can immediately note that the W$_0$ values are shockingly similar across the redshift bins. Only the very faintest lines such as Ti II 1910 and Ti II 3067, which would be most affected by reduced sample size, show significant variation.

Further, when we inspect the cut-outs shown at the bottom of Figure \ref{z-bins} we can see that all plotted lines are remarkable similar across the three redshift bins. These results strongly suggest that when selected by the presence of Mg II absorption there is little to no redshift evolution in the strengths of the considered absorption features.

\section{Discussion}

In this section we first detail the line identifications we have made that proved particularly challenging. We then consider how the detectability of the emission lines depends on both \voff\ and the number of absorbed spectra used to construct the composite.

\subsection{Challenging Line Identifications}
While examining our full sample of absorption lines, we encountered a number of challenges in identification. Here we will document the natures of these challenges, as well as explain our reasoning for assigning a given identification.

Firstly, we considered two possible identifications from M03 for the line we observe at 1039$\textrm{\AA}$, either O I 1039.2304$\textrm{\AA}$ or Fe II 1039.3128$\textrm{\AA}$. As we observe another O I line at 1302 we can consider the relative strength of the two lines, which is expected to be roughly 5:1, as the line identified at 1039 is stronger than this diagnostic would suggest, we adopt the Fe II identification.

Concerning the line we observe at 1122$\textrm{\AA}$, we considered three possible identifications from M03, C I 1122.2597$\textrm{\AA}$, C I 1122.437$\textrm{\AA}$ and Fe III 1122.5183$\textrm{\AA}$. Considering other lines of C I we would expect to detect, we can note the absence of the C I 1129 line, which should be 50\% stronger than C I 1122 if present. As such, we adopt the Fe III identification.

Finally, we can consider the lines we detect at 1454$\textrm{\AA}$ and 1709$\textrm{\AA}$. In both cases we considered two possible identifications, a Ni II line listed in M03, but not found on NIST (1454.842$\textrm{\AA}$ and 1709.6042$\textrm{\AA}$ respectively), and an Fe II line (1454.838$\textrm{\AA}$ and 1709.5543$\textrm{\AA}$ respectively). Despite reservations due to the inability to find the Ni II lines on NIST, we ultimately choose to adopt these identifications. Notably the Ni II 1454$\textrm{\AA}$ line is not present in the finding list for interstellar lines in QSO spectra from \citet[][their Table A2]{York2006}, however the Ni II 1709$\textrm{\AA}$ line is present.

\subsection{Examining Emission Line Detection}
As the emission features we detect in our composite have not been observed to the same signal-to-noise ratio in previously constructed composite spectra, we consider their observation worthy of additional investigation. Firstly, to demonstrate definitively that these emission lines are a feature of the intervening absorption systems, we will analyze composites constructed using only high \voff\ systems. Then we will consider how the detectability of these emission features scales with the number of \Mgii\ absorbers used in the construction of the composite.

When selecting our sample, we choose to use any absorber with \voff\ $ > 3{,}500$ \kms, a fairly canonical value for the separation of intervening and associated absorption systems \citep[e.g.][]{York2006,ZhuMenard2013,Khare2014,Chen2020,Abhijeet2021}. However, this value was commonly thought to be larger in past studies \citep[see discussion in][]{ShenMenard-OII-StarForm}, and even considering the present value there are environments, such as large galaxy groups, in which emission not arising from the absorbing gas could be co-spatial with it.

To verify that the emission lines we detect are truly associated with the absorption line systems, and not the QSO host galaxy, or any nearby group member, we construct two sub-samples, one composed of systems with \voff\ $ > 50{,}000$ \kms, and a second composed of systems with \voff\ $ > 100{,}000$ \kms.

These specific thresholds are chosen arbitrarily, and the resulting sub-samples contain 56.4\% (N = 134{,}610) and 22.2\% (N = 53{,}060) of the full sample respectively. Considering the \voff\ $ > 50{,}000$ \kms\ sub-sample, we can note that the median background QSO redshift in this sample is z$_{\textrm{\tiny QSO}}$ = 2.324, and the median absorption system redshift is z$_{\textrm{\tiny ABS}}$ = 1.244. In a $\Lambda\textrm{CDM}$ cosmology with H$_0$ = 69.6  ${\rm km\,s}^{-1} {\rm Mpc}^{-1}$, $\Omega_M$ = 0.286 and $\Omega_\Lambda$ = 0.714, these redshift values correspond to a difference in comoving radial distance of $\sim 1800$ Mpc, a distance far too great for any causal effect between the background QSO and foreground absorber \citep[][]{CosmoCalc,CosmoCalcSource}. This difference in comoving radial distance determined by the median QSO and absorber system redshift values increases to $\sim 2666$ Mpc for the  \voff\ $ > 100{,}000$ \kms\ sub-sample.

\begin{figure}[ht!]
\epsscale{1.16}
\plotone{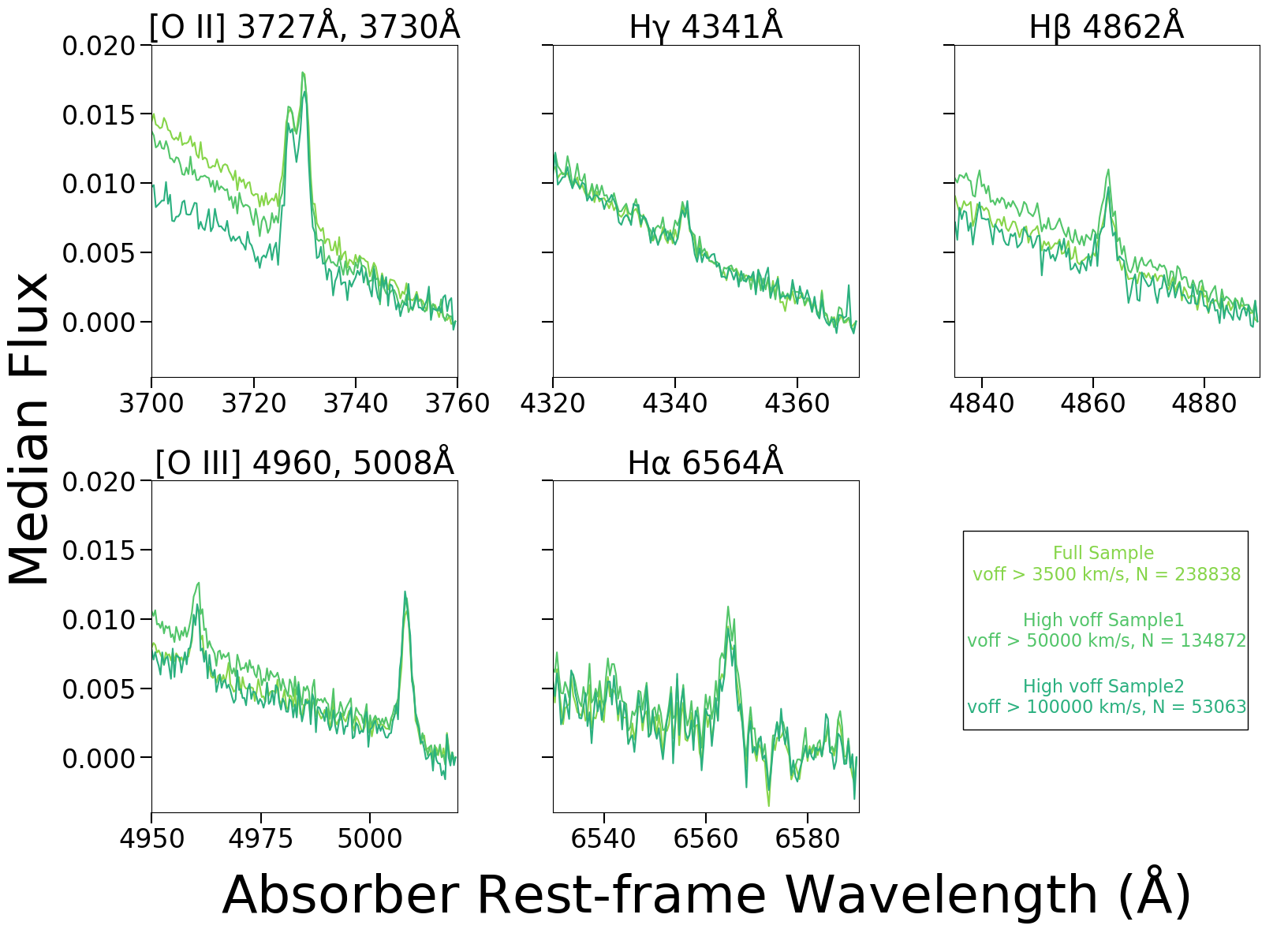}
\caption{The regions of the composite spectrum featuring emission features. Composites are plotted for the full sample as well as two high \voff\ sub-samples as indicated in the legend in the bottom right. Note that the y-axis has been arbitrarily rescaled so as to enable comparison between the different composites.}
\label{Voff-bins}
\end{figure}

Using these sub-samples we then construct composite spectra and plot the emission line regions in Figure \ref{Voff-bins}. We can immediately note that the emission features are nearly identical in all three composites, clearly confirming that these features are indeed associated with the absorption systems.

\begin{figure}[ht!]
\epsscale{1.16}
\plotone{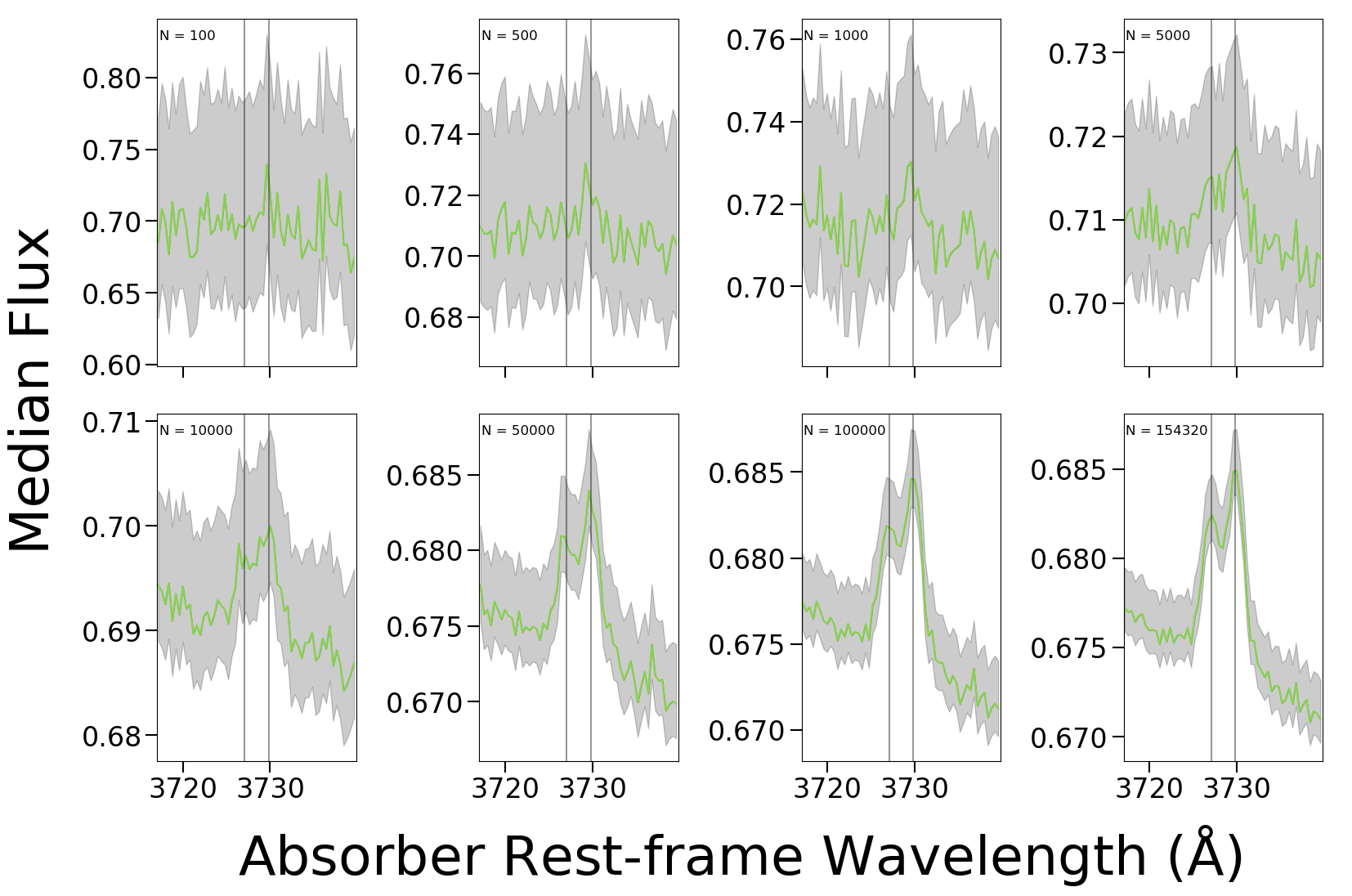}
\caption{The [O II] doublet, 3727$\textrm{\AA}$, 3729$\textrm{\AA}$, as resolved in composite spectra constructed with a varying number of absorption systems. The number of systems contributing to each composite is shown in the upper left hand of each subplot. The central green line is the median composite spectrum, and the gray bands show the error in the composite measurement. Note that the y-scale changes significantly between panels.}
\label{OII-Nbins}
\end{figure}

Next, we consider the detectability of these emission features. Specifically we will consider the [O II] doublet at 3727$\textrm{\AA}$, 3730$\textrm{\AA}$, as it is both the strongest emission feature by equivalent width in the reduced composite, and it is covered by the largest fraction of the sample, as can be seen in Figure \ref{WaveCov}. All absorbers at redshifts $0.3 < z{_{\textrm{\tiny ABS}}} < 1.6$ when shifted into their restframe will cover the [O II] doublet; the total number of such systems is 154{,}138, $\sim 64.5\%$ of our total sample. From this we draw random samples of smaller numbers of absorption systems and compute composite spectra. These composites are collectively shown in Figure \ref{OII-Nbins}. We observe that the [O II] emission doublet is indistinguishable from the noise in composites having less than N = 5{,}000 spectra, and that the doublet only begins to be well-resolved with N $\sim$ 10{,}000 spectra. Here we can clearly see the increase in signal-to-noise ratio gained by constructing an absorber composite spectrum with such a large sample.


Having demonstrated both that the emission features we detect are related to the absorption systems, and that their extremely low signal likely made them undetectable in many previously constructed composites, we consider our detection to be extremely robust.

\section{Conclusion}
In this paper we have described the construction and analysis of a composite spectrum of QSO absorption line systems using a catalog of \Mgii\ absorbers identified in spectra from the second data release (DR2) of DESI. We have outlined the methods by which we constructed our sample of 239{,}174 absorbers, considering only those QSO spectra with a single identified \Mgii\ absorber at \voff\ $ > 3{,}500$ \kms. We further construct a control sample of QSO spectra with no detected \Mgii\ absorbers, by matching to our absorbed sample on the basis of QSO redshift and signal-to-noise ratio

For both our absorbed QSO and control QSO sample, we have calculated median spectra in the restframe of the detected absorption systems. We have outlined the necessary procedures including the resampling of flux values on a common wavelength grid from [1030$\textrm{\AA}$, 7500$\textrm{\AA}$] with a 0.4$\textrm{\AA}$ spacing and the normalization of individual spectra using the average flux in regions adjacent to the \Mgii\ absorption doublet. The resulting composite spectra are presented in Figure \ref{Initial-Comp}.

We have then detailed the methods by which we searched for absorption features in the absorber sample composite spectrum, finding a total of 72 absorption lines from 26 atomic species. We also identify 7 emission lines, associated with the [O II], H$\gamma$, H$\beta$, [O III] and H$\alpha$ transitions. Some of these features, particularly [O II], have been observed and utilized in previous studies of absorber composite spectra \citep[e.g.][]{Wild2007-CaIIComposites,Menard2011,ZhuMenard2013,Khare2014,Ravi2018-OIIEmission}, whereas others (H$\beta$, H$\gamma$) have only been previously observed in QSO-galaxy pairs \citep[e.g.][]{York2012-Halpha,Straka2013-EmissionLines,Straka2015-EmissionLines}. Here we produce the highest signal-to-noise ratio measurement of these lines in an absorber composite spectrum presently available. 

In order to fit the lines we observe, we first estimate a continuum for our absorbed sample composite spectrum using the control sample composite spectrum and a cubic b-spline fit. We have detailed the methods by which we fit the resulting reduced absorber composite, and the results of this line-fitting procedure are presented in our full line atlas (Tables ~\ref{table:abslineatlas} and ~\ref{table:emlineatlas}).

We have additionally constructed composite spectra using subsamples of our absorber population, specifically dividing our sample into three bins by both their \EWFL\ and z$_{\textrm{\tiny ABS}}$ values.

When considering the \EWFL\ subsamples we find that the W$_0$ of all selected lines is highest in the composite spectrum constructed using the largest \EWFL\ values, and that in generally the strengths of all selected lines are positively correlated with \EWFL. However, we do note that this positive correlation was weaker for high ionization species such as C IV and Si IV, reinforcing that they may exist primarily in separate gas phases from Mg II. Additionally, we find that the W$_0$ values of the Na I doublet are much stronger in the highest \EWFL\ bin, suggesting that the dense shielded gas necessary to host Na I may only be found in sightlines that also intercept strong Mg II absorption, such as could be produced by multiple intercepted clouds.

Examining the z$_{\textrm{\tiny ABS}}$ subsamples we find that for absorption lines in the region from 1636 $\textrm{\AA}$ to 3638 $\textrm{\AA}$ their W$_0$ values remain similar across the three selected redshift bins. These results then indicate that from redshifts z$_{\textrm{\tiny ABS}}$ = 0.287 to z$_{\textrm{\tiny ABS}}$ = 2.505 the strengths of the examined absorption lines, when selected by the presence of Mg II absorption, do not significantly vary. Further analysis and larger samples should allow for finer determinations when considering these subsample composites.

We further describe the process by which we derive line identifications, relying primarily on \citet{Morton03} and \citet{NIST}. We have noted how we made identifications for some challenging lines. Finally, we carefully verified that the emission lines we observed in our absorber composite are genuinely at the same redshift as the gas associated with the absorbers, by constructing composites only at very high \voff. We then also examined the increase in signal-to-noise of the [O II] doublet with increasing sample size, demonstrating that the emission features present in this composite would not have been readily detectable in many previous absorber composite studies. 

The details presented in the paper provide the most complete study of the composite spectrum of QSO absorption line systems in the wavelength range [1030$\textrm{\AA}$, 7500$\textrm{\AA}$] to date, and should aid in further studies of the physical natures of these absorbers. Furthermore, our analysis here provides an opportunity to better model absorption line systems, and understand their impact on studies of the \LyA\ forest, and other studies of the spectra of QSOs. In future work we intend to further examine the subsamples composites formed when binning absorbers based on their \EWFL\ and z$_{\textrm{\tiny ABS}}$ values, and to construct similar composites for the DESI DR3 sample, when made available.

\begin{acknowledgments}
We would like to thank the anonymous referee for their comments and suggestions that have greatly improved the
content and usefulness of this paper.

LN and ADM were supported by the U.S.\ Department of Energy, Office of Science, Office of High Energy Physics, under Award Number DE-SC0019022.

This material is based upon work supported by the U.S. Department of Energy (DOE), Office of Science, Office of High-Energy Physics, under Contract No. DE–AC02–05CH11231, and by the National Energy Research Scientific Computing Center, a DOE Office of Science User Facility under the same contract. Additional support for DESI was provided by the U.S. National Science Foundation (NSF), Division of Astronomical Sciences under Contract No. AST-0950945 to the NSF’s National Optical-Infrared Astronomy Research Laboratory; the Science and Technology Facilities Council of the United Kingdom; the Gordon and Betty Moore Foundation; the Heising-Simons Foundation; the French Alternative Energies and Atomic Energy Commission (CEA); the National Council of Humanities, Science and Technology of Mexico (CONAHCYT); the Ministry of Science, Innovation and Universities of Spain (MICIU/AEI/10.13039/501100011033), and by the DESI Member Institutions: \url{https://www.desi.lbl.gov/collaborating-institutions}. Any opinions, findings, and conclusions or recommendations expressed in this material are those of the author(s) and do not necessarily reflect the views of the U. S. National Science Foundation, the U. S. Department of Energy, or any of the listed funding agencies.

The authors are honored to be permitted to conduct scientific research on I'oligam Du'ag (Kitt Peak), a mountain with particular significance to the Tohono O’odham Nation.
\end{acknowledgments}

The data used to construct the Figures in this paper are available on Zenodo at \dataset[DOI: 10.5281/zenodo.17497237]{https://doi.org/10.5281/zenodo.17497238}

\bibliographystyle{yahapj}
\bibliography{AbsEXT.bib}

\startlongtable
\begin{deluxetable*}{cccc|cc}
\label{table:abslineatlas}
\tabletypesize{\footnotesize}
\tablecaption{Absorption Lines Observed in Absorber Composite}
\tablecolumns{5}
\tablewidth{0pt}
\tablehead{
\colhead{Vacuum} & \colhead{} & \colhead{Oscillator} & \colhead{} &  \colhead{Observed} & \cr
\colhead{Wavelength (\AA)} & \colhead{Species} & \colhead{Strength ($f_{ik}$)} & \colhead{Reference} & \colhead{Wavelength (\AA)} & \colhead{$W_0$ (\AA)}
}
\startdata
1031.9268 & O VI  & 1.33E-1 & NIST & 1031.882 & ${1.511}^{+0.472}_{-0.236}$ \\
1036.335 & C II  & 1.19E-1 & NIST & 1036.359 & ${1.236}^{+0.568}_{-0.128}$ \\
1037.6154 & O VI  & 6.60E-2 & NIST & 1037.584 & ${0.980}^{+0.272}_{-0.980}$ \\
1039.3128 & Fe II  & 1.05E-4 & M03 & 1039.175 & ${0.754}^{+0.194}_{-0.164}$ \\
1063.020 & Fe II  & 1.4E-2 & NIST & 1063.136 & ${0.946}^{+0.111}_{-0.117}$ \\
1081.872 & Fe II  & 1.3E-2 & NIST & 1081.793 & ${0.619}^{+0.155}_{-0.073}$ \\
1083.990 & N II  & 1.11E-1 & NIST & 1083.977 & ${1.004}^{+0.038}_{-0.079}$ \\
1096.871 & Fe II & 3.3E-2 & NIST & 1096.795 & ${0.503}^{+0.053}_{-0.044}$ \\
1122.524 & Fe III  & 5.44E-2 & M03 & 1122.283 & ${0.612}^{+0.033}_{-0.033}$ \\
1125.442 & Fe II  & 1.6E-2 & NIST & 1125.492 & ${0.360}^{+0.061}_{-0.043}$ \\
1134.520 & N I$^{ a}$  & 1.46E-2/2.87E-2/4.16E-2 & NIST & 1134.571 & ${0.728}^{+0.028}_{-0.027}$ \\
1143.220 & Fe II & 1.9E-2 & NIST & 1143.207 & ${0.393}^{+0.039}_{-0.035}$ \\
1144.926 & Fe II & 8.3E-2 & NIST & 1144.963 & ${0.676}^{+0.022}_{-0.022}$ \\
1190.42 & Si II & 2.77E-1 & NIST & 1190.398 & ${1.213}^{+0.010}_{-0.010}$ \\
1193.28 & Si II & 5.75E-1 & NIST & 1193.295 & ${1.290}^{+0.009}_{-0.009}$ \\
1200.161 & N I$^b$ & 1.32E-1/8.69E-2/4.32E-2 & NIST & 1200.074 & ${1.002}^{+0.013}_{-0.015}$ \\
1206.51 & Si III & 1.67E0 & NIST & 1206.506 & ${1.443}^{+0.007}_{-0.009}$ \\
1215.673 & Ly$\alpha$ & 4.164E-1 & NIST & 1215.653 & ${9.750}^{+0.042}_{-0.056}$ \\
1238.804 & N V & 1.56E-1 & NIST & 1238.903 & ${0.506}^{+0.032}_{-0.036}$ \\
1242.795 & N V & 7.80E-2 & NIST & 1242.767 & ${0.335}^{+0.041}_{-0.028}$ \\
1250.5834 & S II & 6.02E-3 & NIST & 1250.538 & ${0.303}^{+0.033}_{-0.025}$ \\
1253.813 & S II & 1.21E-2 & NIST & 1253.783 & ${0.410}^{+0.021}_{-0.021}$ \\
1260.42 & Si II & 1.22E0 & NIST & 1260.398 & ${1.430}^{+0.007}_{-0.007}$ \\
1302.168 & O I & 5.20E-2 & NIST & 1302.173 & ${1.126}^{+0.007}_{-0.006}$ \\
1304.37 & Si II & 9.28E-2 & NIST & 1304.369 & ${0.997}^{+0.008}_{-0.007}$ \\
1317.22 & Ni II$^c$ & - & NIST & 1317.245 & ${0.280}^{+0.029}_{-0.025}$ \\
1334.5323 & C II & 1.29E-1 & NIST & 1334.547 & ${1.468}^{+0.005}_{-0.005}$ \\
1370.14 & Ni II & - & NIST & 1370.147 & ${0.290}^{+0.021}_{-0.018}$ \\
1393.76 & Si IV & 5.13E-1 & NIST & 1393.762 & ${1.178}^{+0.005}_{-0.004}$ \\
1402.77 & Si IV & 2.55E-1 & NIST & 1402.770 & ${0.972}^{+0.005}_{-0.005}$ \\
1454.842 & Ni II & 1.9E-3 & M03 & 1454.815 & ${0.165}^{+0.022}_{-0.021}$ \\
1526.72 & Si II & 1.33E-1 & NIST & 1526.708 & ${1.152}^{+0.003}_{-0.003}$ \\
1548.2043 & C IV & 1.90E-1  & NIST & 1548.210 & ${1.501}^{+0.003}_{-0.003}$ \\
1550.784 & C IV & 9.52E-2 & NIST & 1550.769 & ${1.311}^{+0.003}_{-0.003}$ \\
1560.3083 & C I & 7.16E-2 & NIST & 1560.345 & ${0.260}^{+0.012}_{-0.015}$ \\
1608.4508 & Fe II & 5.91E-2 & NIST & 1608.451 & ${0.771}^{+0.004}_{-0.005}$ \\
1611.197 & Fe II & 1.4E-3 & NIST & 1611.243 & ${0.191}^{+0.034}_{-0.021}$ \\
1656.9282 & C I & 1.43E-1 & NIST & 1657.010 & ${0.305}^{+0.013}_{-0.013}$ \\
1670.7867 & Al II & 1.77E0 & NIST & 1670.788 & ${1.188}^{+0.003}_{-0.003}$ \\
1709.6042 & Ni II & 2.6E-3 & M03 & 1709.577 & ${0.275}^{+0.011}_{-0.010}$ \\
1741.55 & Ni II & - & NIST & 1741.556 & ${0.306}^{+0.009}_{-0.009}$ \\
1751.9157 & Ni II & - & M03 & 1751.959 & ${0.231}^{+0.011}_{-0.011}$ \\
1808.00 & Si II & 2.49E-3 & NIST & 1808.002 & ${0.389}^{+0.006}_{-0.006}$ \\
1854.7164 & Al III & 5.61E-1 & NIST & 1854.720 & ${0.729}^{+0.003}_{-0.003}$ \\
1862.7895 & Al III & 2.79E-1 & NIST & 1862.798 & ${0.557}^{+0.004}_{-0.004}$ \\
1910.7831 & Ti II$^d$ & 1.04E-1/9.8E-2 & NIST & 1910.690 & ${0.141}^{+0.016}_{-0.016}$ \\
2026.295 & Zn II/Cr II/Mg I$^e$ & 5.01E-1/1.30E-3/1.13E-1 & NIST & 2026.214 & ${0.333}^{+0.006}_{-0.006}$ \\
2056.2568 & Cr II & 1.03E-1 & NIST & 2056.243 & ${0.255}^{+0.007}_{-0.007}$ \\
2062.4483 & Cr II/Zn II$^f$ & 7.59E-2/2.46E-1 & NIST & 2062.384 & ${0.299}^{+0.006}_{-0.006}$ \\
2066.1639 & Cr II & 5.12E-2 & NIST & 2066.163 & ${0.184}^{+0.010}_{-0.010}$ \\
2249.8768 & Fe II & 1.82E-3 & M03 & 2249.877 & ${0.266}^{+0.006}_{-0.006}$ \\
2260.7805 & Fe II & 2.44E-3 & M03 & 2260.780 & ${0.323}^{+0.005}_{-0.005}$ \\
2344.2139 & Fe II & 1.14E-1 & M03 & 2344.212 & ${1.083}^{+0.002}_{-0.002}$ \\
2374.4612 & Fe II & 3.13E-2 & M03 & 2374.463 & ${0.794}^{+0.002}_{-0.002}$ \\
2382.7652 & Fe II & 3.20E-1 & M03 & 2382.766 & ${1.324}^{+0.002}_{-0.002}$ \\
2576.8752 & Mn II & 3.58E-1 & NIST & 2576.883 & ${0.347}^{+0.004}_{-0.004}$ \\
2586.6500 & Fe II & 6.91E-2 & M03 & 2586.648 & ${1.028}^{+0.002}_{-0.002}$ \\
2594.4966 & Mn II & 2.79E-1 & NIST & 2594.503 & ${0.323}^{+0.004}_{-0.004}$ \\
2600.1729 & Fe II & 2.39E-1 & M03 & 2600.173 & ${1.334}^{+0.001}_{-0.002}$ \\
2606.4587 & Mn II & 1.96E-1 & NIST & 2606.465 & ${0.275}^{+0.005}_{-0.004}$ \\
2796.352 & Mg II & 6.08E-1 & NIST & 2796.354 & ${1.873}^{+0.001}_{-0.001}$ \\
2803.530 & Mg II & 3.03E-1 & NIST & 2803.533 & ${1.725}^{+0.001}_{-0.001}$ \\
2852.965 & Mg I & 1.80E0 & NIST & 2852.962 & ${0.742}^{+0.002}_{-0.002}$ \\
3067.2379 & Ti II & 4.89E-2 & M03 & 3066.973 & ${0.110}^{+0.016}_{-0.018}$ \\
3073.8633 & Ti II & 1.21E-1 & M03 & 3073.780 & ${0.167}^{+0.009}_{-0.009}$ \\
3230.1221 & Ti II & 6.87E-2 & M03 & 3230.042 & ${0.136}^{+0.012}_{-0.015}$ \\
3242.9184 & Ti II & 2.32E-1 & M03 & 3242.933 & ${0.211}^{+0.007}_{-0.007}$ \\
3384.7304 & Ti II & 3.58E-1 & M03 & 3384.720 & ${0.291}^{+0.007}_{-0.006}$ \\
3934.777 & Ca II & 6.82E-1 & NIST & 3934.763 & ${0.552}^{+0.005}_{-0.005}$ \\
3969.591 & Ca II & 3.3E-1 & NIST & 3969.567 & ${0.413}^{+0.006}_{-0.006}$ \\
5891.5832 & Na I & 6.41E-1 & NIST & 5891.522 & ${0.467}^{+0.031}_{-0.048}$ \\
5897.5581 & Na I & 3.20E-1 & NIST & 5897.522 & ${0.321}^{+0.075}_{-0.038}$ \\
\hline
\enddata
\end{deluxetable*}

\onecolumngrid

\footnotesize{$^a$ Blend of N I lines: 1134.165$\,\textrm{\AA}$, 1134.415$\,\textrm{\AA}$, 1134.980$\,\textrm{\AA}$.}

\footnotesize{$^b$ Blend of N I lines: 1199.550$\,\textrm{\AA}$, 1200.223$\,\textrm{\AA}$, 1200.710$\,\textrm{\AA}$.}

\footnotesize{$^c$ Note that oscillator strength values are not available on NIST for many of the Ni II lines we observe.

\footnotesize{$^d$ Blend of Ti II lines: 1910.6123$\,\textrm{\AA}$, 1910.9538$\,\textrm{\AA}$.}

\footnotesize{$^e$ Blend of Zn II 2026.1371$\,\textrm{\AA}$, Cr II 2026.271$\,\textrm{\AA}$, Mg I 2026.477$\,\textrm{\AA}$.}

\footnotesize{$^f$ Blend of Cr II 2062.2360$\,\textrm{\AA}$, Zn II 2062.6604$\,\textrm{\AA}$.}

\clearpage

\startlongtable
\begin{deluxetable*}{cccc|cc}
\label{table:emlineatlas}
\tabletypesize{\footnotesize}
\tablecaption{Emission Lines Observed in Absorber Composite}
\tablecolumns{5}
\tablewidth{0pt}
\tablehead{
\colhead{Vacuum} & \colhead{} & \colhead{Oscillator} & \colhead{} & \colhead{Observed} & \cr
\colhead{Wavelength (\AA)} & \colhead{Species} & \colhead{Strength ($f_{ik}$)} & \colhead{Reference} & \colhead{Wavelength (\AA)} & \colhead{$W_0$ (\AA)}
}
\startdata
3727.092 & [O II] & 3.31E-13 & NIST & 3727.112 & ${0.251}^{+0.019}_{-0.017}$ \\
3729.875 & [O II] & 8.95E-14 & NIST & 3729.891 & ${0.304}^{+0.013}_{-0.016}$ \\
4341.692 & H$\gamma$ & 4.469E-2 & NIST & 4341.485 & ${0.177}^{+0.034}_{-0.031}$ \\
4862.71 & H$\beta$ & 1.194E-1 & NIST & 4862.658 & ${0.219}^{+0.018}_{-0.034}$ \\
4960.295 & [O III] & 2.81E-14 & NIST & 4960.259 & ${0.260}^{+0.024}_{-0.022}$ \\
5008.240 & [O III] & 1.32E-13 & NIST & 5008.334 & ${0.428}^{+0.015}_{-0.015}$ \\
6564.60 & H$\alpha$ & 6.411E-1 & NIST & 6564.704 & ${0.444}^{+0.048}_{-0.053}$ \\
\hline
\enddata
\end{deluxetable*}

\onecolumngrid

\end{document}